\documentclass[12pt]{article}
\usepackage{a4}
\usepackage{amssymb}
\usepackage{epsf}
\usepackage{cite}
\def\Bbb{\mathbb}
\def\BZ{\Bbb Z} \def\BR{\Bbb R}
\def\BC{\Bbb C} \def\BP{\Bbb P}

\catcode`@=11 \@addtoreset{equation}{section} \catcode`@=12

\begin{document}
\begin{titlepage}
\noindent
\phantom \hfill
{\tt hep-th/0010196}\\
{\tt IMSc/2000/10/57}~~~
\hfill October, 2000
\vfill
\begin{center}
{\Large \bf D-branes, Exceptional Sheaves and Quivers
on Calabi-Yau manifolds: From Mukai to McKay} \\[1cm]
Suresh Govindarajan\\ 
{\em Department of Physics, Indian Institute of Technology, Madras,\\
Chennai 600 036, India }\\
{\tt Email: suresh@chaos.iitm.ernet.in} \\[10pt]
and \\[10pt]
T. Jayaraman\\ 
{\em The Institute of Mathematical Sciences, \\ Chennai 600 113,
India}\\
{\tt Email: jayaram@imsc.ernet.in}\\[10pt]
\end{center}
\vfill
\begin{abstract}
We present a method based on mutations of helices
which leads to the  construction (in the large volume limit)
of exceptional coherent sheaves associated with
the $(\sum_al_a=0)$ orbits in Gepner models. This is explicitly verified
for a few examples including some cases where the ambient weighted
projective space has singularities not inherited by the Calabi-Yau
hypersurface.  The method is based on two conjectures
which lead to the analog,in the general case, of the Beilinson quiver for
$\BP^n$. We discuss how one recovers the McKay quiver using the
gauged linear sigma model (GLSM) near the orbifold or Gepner point 
in K\"ahler moduli space.
\end{abstract}
\vfill
\end{titlepage}

\section{Introduction}

There has been significant progress in the
study of D-branes on Calabi-Yau manifolds   
in the recent past following the construction of
D-branes at Gepner points in Calabi-Yau moduli space.  
Among the significant results have been the description of the A- and
B-type boundary states, corresponding to D-branes,
in Gepner models and their transformation, under monodromy
action as one moves  in K\"ahler moduli space, 
to particular large-volume D-brane
configurations\cite{RS,quintic,diacgom,dgepner,stt,scheid}.
But once explicit results from the Gepner points are available,
one of the important issues
is to understand how these descriptions of D-branes at Gepner points can
be recovered in a reasonably straightforward and algorithmic manner from
the description of D-branes on large-volume CY manifolds. More
generally, it would be useful to derive an understanding of the D-brane
spectrum and other related properties including aspects of their
world-volume theories in all possible ``phases'' of CY manifolds.
While the ability to perform explicit computations is the reason to study
some points in the K\"ahler moduli space, like the Gepner points, 
the large-volume ``phase'' of CY manifolds provides a 
clearer geometric picture which is useful to understand in detail. As is
well known in the closed string case, 
all ``phases'' can be reached by 
choosing appropriate values of the Fayet-Iliopoulos (FI) terms in the
in the gauged linear sigma model (GLSM) \cite{wittenphases}. 
Thus the GLSM seems to be the logical 
starting point for the study of D-branes in the different phases of the 
CY manifolds. Of course, in the case of D-branes, we will have to
understand the problem of marginal stability, whereby D-branes
present at one point in K\"ahler moduli space may decay into other
branes as the moduli are varied. But nevertheless one may hope to
address even these issues by some means through the use of the GLSM. 
In this paper it is this point of view, of trying to connect the
behaviour of the D-brane spectrum in all the different phases by using
the GLSM description of D-branes, that we will pursue.  

Various aspects of such a connection between the D-brane spectrum in
different phases, and in particular at the 
the large volume limit and the Gepner point have already begun emerging
from the work of \cite{diacgom,topics,dfr1,dfr2,DD,dougtalk}. 
Among the most important result of these
studies has been the uncovering of the special role of classes of
exceptional bundles (or more generally exceptional sheaves)
on CY manifolds. These D-brane configurations (which have no moduli)
appear to be the building blocks of other brane configurations with
moduli. The D-brane configurations which are seen at the Gepner point by
the Recknagel-Schomerus\cite{RS} construction are all in this class.
The $\sum_a l_a=0$  orbit of B-type boundary states are all 
related to exceptional bundles in the large volume limit by monodromy 
transformations. The other B-type boundary states seen at the Gepner 
point with $\sum_a l_a\neq 0$ are clearly bound
states of these exceptional objects\cite{dfr2,DD}.

The study of the case of  
the $\BC^3/\BZ_3$ and its blowup to $\BP^2$ first showed a natural connection
between the fractionally charged branes (or equivalently
wrapped branes pinned to the orbifold point in $\BC^3$) and a
particular class of exceptional bundles on $\BP^2$\cite{diacgom,dfr2}.
Whereas the branes that exist at the orbifold point can be described by a
quiver gauge theory, based on the McKay quiver\footnote{for earlier 
related work on quiver gauge theories
and D-branes on orbifolds which are directly related to our context,
 see \cite{dgm,inverse}.}, the corresponding
bundles on the blowup are related to a truncated version of
this quiver, namely the Beilinson quiver. Moreover the
Beilinson quiver can be seen to be 
equivalent to the monad construction for bundles on $\BP^2$. 
Further, any sheaf on $\BP^2$ can be written as a cohomology of a complex
involving the bundles that are the large-volume monodromy transforms of
the fractional branes in the orbifold limit. This permits the description 
of D-branes in
the large volume limit in terms of the numbers naturally associated
with the labelling of
charges of D-branes in the orbifold limit. This relationship immediately
provides a handle on the question of marginal stability. Whereas the
charges of stable D-branes at the orbifold point
(in the orbifold basis) have to be either all positive or all negative, the
bundles in the large volume have no such restriction. Clearly
the candidates for decay (in the passage) from the large-volume limit
to the orbifold point, are those branes which have a combination of
negative and positive charges in the orbifold basis.
 
The extension of this procedure to the study of the Landau-Ginzburg(LG)
theory associated to the Gepner point of a CY manifold 
is a logical one. One may regard
the LG theory as an orbifold of the type $\BC^n/\Gamma$ if one disregards
the superpotential. Douglas and Diaconescu\cite{DD} (see also
\cite{diacgom} for related observations) 
considered only the so-called tautological line bundles on the
orbifold $\BC^n/\Gamma$ (this is equivalent to considering only the states
associated with the vector representation of $Sl(n,\BC)$). Then 
by using the 
inverse procedure of reconstructing a toric variety from the quiver data
associated with the orbifold (using the methods of \cite{dgm,inverse}),
 that one can construct the corresponding line bundles,
$\{R_i\}$ on the
space obtained by the resolution of the singularities of $\BC^n/\Gamma$.
One then constructs a set of vector bundles $\{S_i\}$ on the resolved
space which are dual (in the sense of K-theory) to the $\{R_i\}$.
The restriction of these $\{S_i\}$ 
to the CY hypersurface provides exactly the set of
exceptional vector bundles $\{V_i\}$ associated with the $\sum_al_a=0$ orbit 
of D-branes at the Gepner point. More precisely, it was verified in
\cite{DD} that the
D-brane charges associated with the $\{V_i\}$ were exactly the same as those
obtained by monodromy transformation of the D-brane charges of the
$\sum_al_a=0$ orbit at the Gepner point. 

However the methods used by Douglas and Diaconescu to derive 
this remarkable result have two obvious limitations.
First, the inverse procedure of reconstructing the toric variety that
corresponds to resolving the singularities of the orbifold at the Gepner
point is a cumbersome one. More importantly, it is not clear that 
there is a canonically-defined procedure for all resolutions of the
singularities. Generically the CY hypersurface
may not intersect some of the singularities and it is not clear 
how one may deal with such situations.

It is the attempt of this paper to show that in the frame-work of the
GLSM there is a method to construct a set of objects analogous to the
fundamental objects of the
Douglas-Diaconescu construction\footnote{While this construction draws
inspiration in an important way from the mathematical 
work of Ito and Nakajima \cite{Ito-Nak}, the authors' use of this work goes
considerably beyond its original form. The reviews by Miles Reid of the
Mckay correspondence
\cite{reid1,reid2} are also useful in this context.}, namely the $\{R_i\}$,
$\{S_i\}$, and
the $\{V_i\}$. The $\{V_i\}$ that we construct are identical to the set
constructed by Diaconescu and Douglas in the examples they consider. 
The objects $\{R_i\}$ and $\{S_i\}$ are
similar to theirs in the sense that the $\{R_i\}$ are a set of line bundles
and that they are dual to the $\{S_i\}$ in the sense
of intersection theory. Further the restriction of $\{S_i\}$ to the CY
hypersurface produces exactly the $\{V_i\}$.
Furthermore we construct the $\{R_i\}$ in a canonical manner
using the GLSM in the neighbourhood of the large volume limit and show that the
$\{S_i\}$ are defined by a consistent mathematical procedure that has a
nice physical interpretation. The construction of the $\{V_i\}$ is then a
straightforward matter. In a number of examples with one and two K\"ahler
moduli we verify our claims. From the $\{R_i\}$ that we have constructed we
can in fact reconstruct the quiver that is the analogue of the Beilinson
quiver in the general case. We take a further step and argue that the 
presence of the so-called $p$-field in the GLSM allows us to reconstruct 
the McKay quiver that is associated to the orbifold LG theory, 
(which indeed may or may not have a Gepner construction). 

In the next subsection we will sketch a more
detailed overview of our method and results before we flood the reader
with more technical details.

\subsection{From Mukai to McKay}

The natural starting point for the construction of rigid D-branes on CY
manifolds is the 6-brane. Clearly there are no moduli in this case. In
more mathematical terminology this is associated to the line bundle of
degree zero on the CY manifold $M$, denoted by ${\cal O}_M$. We can
obtain 
this line bundle by restricting the line bundle ${\cal O}$
on the ambient variety $X $ to the CY. We obtain an
infinite set of other rigid objects $R_i,i\in \BZ$, through the
action of large-volume monodromy. The effect of the shift $B \rightarrow
B + 1$ is implemented by the shift $\theta \rightarrow \theta + 2\pi$ in
the GLSM. This has the effect of tensoring the initial bundle by 
a line bundle\cite{glsm,HIV}. This is physically equivalent to 
turning on D4-brane charges on the 6-brane together with lower brane charges. 

Confining ourselves to the case where $X$ is $\BP^n$ for the
moment, we can restrict our considerations to the set $R_i={\cal O}(i-1)$,
where $i$ runs over $1$ to $(n+1)$. This collection of line bundles
${\cal R}\equiv \{R_i\}$ will be our starting point.
The collection ${\cal R}$ is referred to as the
foundation of a helix for reasons that will be made clear later. 
From the work of Rudakov and others\cite{rudakov} we have a
well-defined mathematical procedure, known as mutations, that enable us
to construct the $\{S_i\}$. This procedure has a nice physical
interpretation as has been demonstrated by Hori, Iqbal and Vafa
\cite{HIV} (see also the earlier paper of Zaslow\cite{zaslow}).
Mutations correspond to brane creation in the mirror of the original
variety $X$. Take a neighbouring pair $(R_i, R_{i+1})$ in the set
${\cal R}$ and consider the corresponding  middle dimensional cycles on
the mirror. In the mirror, a mutation 
corresponds to the creation of a new middle dimensional brane when the
$R_i$ ``crosses'' $R_{i+1}$. The mirror of this new
middle-dimensional brane in the original variety $X$ is referred to as the
mutation of the two $R_i$ that were our starting point. 

A sequence of such mutations enables us to construct all the $\{S_i\}$ up to
a sign that is fixed so that the $\{S_i\}$ are in fact dual to the
$\{R_i\}$.
If the Chern classes of the $\{S_i\}$ are all that we are interested in then
of course this can be obtained by simply examining the inverse of the
Euler matrix of the $\{R_i\}$, as is indeed clear from the work of
Douglas and Diaconescu. On restricting the $\{S_i\}$ to the CY
hypersurface, one obtains the $\{V_i\}$.

For the general case, going beyond $\BP^n$, we formulate two conjectures
that we verify in a variety of one and two K\"ahler modulus examples.\\

\noindent {\bf Conjecture 1}: The large volume monodromy action on 
${\cal O}$ in the ambient variety $X$
produces an exceptional collection which is the foundation
of a  helix of appropriate period $p$:
\begin{equation}
{\cal R}= (R_1 ={\cal O}, R_2, \ldots, R_p)
\end{equation}

\noindent {\bf Conjecture 2}: 
All exceptional bundles and sheaves on $X$ 
can be obtained by the mutations of the helix
${\cal R}$ given in Conjecture 1. In particular, there exists a
mutated helix ${\cal S}=(S_p,\ldots,S_1={\cal O})$ with
$S_i=L^{i-1}(R_i)$, where $L^i$ corresponds to a sequence of $i$
left-mutations.

\noindent {\bf Conjecture 2a}:  All the exceptional bundles on the CY
which correspond to the the $\sum_a l_a =0$ states at the Gepner
point are obtained by the restriction of the $S_i$ to the CY
hypersurface.

Indeed since exceptional collections and helices are known for a wide
range of varieties including Grassmannians, products of projective and
flag varieties etc.( see \cite{rudakov} 
and references therein) it appears that we can carry out this
procedure for a wide range of CY manifolds that can be described as
hypersurfaces in such ambient varieties. (In the cases where there is no
corresponding Gepner construction, we would have to suitably generalise
conjecture 2a).

The corresponding quiver gauge theory can now be written down in a
canonical way. It is interesting that we have circumvented the problem
of determining which resolution of the singularities of the orbifold 
associated with the Gepner point to deal with, since the large-volume
limit of the GLSM provides us with a definite resolution that we must
deal with. However this quiver is the analogue of the Beilinson quiver
and misses the symmetry at the Gepner point that
is due to the orbifold group. Restoring this symmetry corresponds to
going to the corresponding McKay quiver from the analogue of the
Beilinson quiver. We argue that this is done by the $p$-field. It is the
$p$-field that enables the missing link to be restored in the quiver
diagram that then becomes the McKay quiver. The link was missing in the
constructions of the large-volume limit since the $p$-field is set to
zero in that limit. We leave the detailed
argument to a subsequent section of this paper. This rule provides
us, in
principle, a means to write down the quiver gauge theories corresponding
to the world-volume theories of the branes in different phases of the CY
manifold, depending on which of the several $p$-fields are set to zero
or acquire a non-zero vacuum expectation value in the world-sheet
theory. Finally, by using a generalisation of Beilinson's methods for
$\BP^n$, one can in principle construct all sheaves on the ambient
projective space. The restriction of these sheaves to the Calabi-Yau
hypersurface $M$ provides one with a large collection of
sheaves on $M$.

\subsection{Motivation}

We would like to explain the motivation and background that have led to
the results that are described in this paper. The work arose as a result
of trying to understand D-branes (in particular, D-branes with B-type
boundary conditions) in gauged linear sigma models with boundary. 
In particular, the theta term in the GLSM  Lagrangian imposes
non-trivial constraints on the possible boundary conditions. Further,
one needs to add a boundary {\em contact} term in order that 
boundary conditions in the GLSM have a proper large volume (NLSM)
limit\cite{glsm}.
For the case of a six-brane, i.e., the D-brane associated with line
bundle ${\cal O}$, one can see that this contact term ensures that 
the large-volume monodromy is captured properly in the GLSM with
boundary. This, in fact, carries over to the case of D-branes associated
with non-trivial vector bundles. Thus, the importance of
large-volume monodromy is best seen in the GLSM with
boundary\cite{glsmnew}. 

As is well known, the GLSM interpolates between different phases
of the Calabi-Yau manifold.  Even in the simple example of
Calabi-Yau manifolds with one K\"ahler modulus, a key addition in the
GLSM to the
fields which exist in the NLSM limit is the so-called $p$-field whose
vacuum expectation value (vev) plays the role of an order parameter. 
In considerations of the GLSM with boundary, this field again plays an
important role. As we will see, even if one stays in the geometric
phase, the transition from the NLSM to the GLSM provides extra
information about other phases as well provided we account for the
behaviour of the $p$-field.

Mirror symmetry (in its extended form) also plays an important role in
our considerations though they are not quite required for the results.
The large volume limit gets mapped to a particular point in the 
moduli space of complex structures on the mirror. This point is rather
special -- it is the point of {\em maximal unipotent monodromy} 
and is given by the transverse intersection of divisors corresponding to
degeneration of Calabi-Yau manifolds (see \cite{lty} and references
therein for more details). Further, the process of mutations has
been related to brane creation on the mirror\cite{HIV}.

The paper is organised as follows: In section 2, we set up our notation
as well as conventions. In section 3, we define helices and mutations
and then describe our conjectures. As a result, we give a precise method
to obtain the $\sum_a l_a=0$ bundles for Gepner models associated with Calabi-Yau
manifolds.
In section 4, we test our conjectures
in several examples with one and two K\"ahler moduli. We also show how
one deals with singularities (in the ambient weighted projective space)
that are not inherited by Calabi-Yau hypersurfaces. In section 5, we
show how one constructs the generalised Beilinson quiver from the
helices and then argue how the $p$-field enables one to recover the
McKay quiver associated with the LG orbifold/Gepner model. We conclude
in section 6 with some remarks.

\section{Notation and Conventions}

We will be consider two classes of models in this paper. We shall
use the language of the gauged linear sigma model (GLSM)/toric
description  in the discussion of the models

\subsection{The one-parameter models}  

The GLSM consists of
six chiral superfields $\Phi_i$ $(i=0,1,\ldots,5)$ and a single
abelian vector superfield. The bosonic components $\phi_i$ of the chiral
superfields will later be identified with the quasi-homogeneous
coordinates of some weighted projective space.
The Fayet-Iliopoulos parameter is labelled
$t=r+i{\theta\over{2\pi}}$ and is related to the complexified
K\"ahler modulus $(B+iV)$. The chiral superfield $\Phi_0$ will also be
called the $P$-field and plays a special role. The
charge vector is 
$$
q_i=\pmatrix{q_0&q_1&q_2&q_3&q_4&q_5}\quad,
$$
with $q_0=-\sum_{i=1}^5 q_i$. The D-term constraint 
is given by
$$
D = -e^2 \left(\sum_{i=0}^{5} q_i |\phi_i|^2 -r\right)\quad,
$$
In the absence of a superpotential, in the large volume limit, the
space is a non-compact Calabi-Yau
manifold corresponding to the total space of the line bundle ${\cal
O}(q_0)$ over the weighted projective space $\BP^{q_1,q_2,q_3,q_4,q_5}$.

The introduction of a superpotential $W=PG(\Phi)$, where $G$ is a
quasi-homogeneous function (of $\Phi_i$ for $i\neq0$) of degree
$|q_0|$,
gives rise to a compact Calabi-Yau manifold as a hypersurface $M$
given by the equation
$G=0$ in weighted projective space $X\equiv\BP^{q_1,q_2,q_3,q_4,q_5}$
in the NLSM limit ($e^2 \sqrt r \rightarrow \infty$)\footnote{We
shall hereafter refer to the ambient weighted
projective space as $X$ and the compact Calabi-Yau
hypersurface as $M$. We will also use the terms {\em large volume limit}
and the {\em NLSM limit} in an interchangeable manner.}. Note that, in the 
NLSM limit,
the field $p$ is set to zero in the ground state. 

We shall consider four specific examples given by (studied in
\cite{kt,font})
\begin{eqnarray}
\BP^4[5]=\BP^{1,1,1,1,1}[5]:\quad&& G=\phi_1^5 + \phi_2^5 +\phi_3^5 +\phi_4^5
+\phi_5^5 =0\quad, \\
\BP^{1,1,1,1,2}[6]:\quad&& G=\phi_1^6 + \phi_2^6 +\phi_3^6 +\phi_4^6
+\phi_5^3 =0\quad, \\
\BP^{1,1,1,1,4}[8]: \quad&& G=\phi_1^8 + \phi_2^8 +\phi_3^8 +\phi_4^8
+\phi_5^2 =0\quad, \\
\BP^{1,1,1,2,5}[10]:\quad&& G=\phi_1^{10} + \phi_2^{10} +\phi_3^{10}
+\phi_4^5 +\phi_5^2 =0\quad. 
\end{eqnarray}

Weighted projective spaces are typically singular. This is true of
all the examples we consider except for the case of $\BP^4$. For example,
consider $\BP^{1,1,1,1,2}$. In the chart $\phi_5=1$, there is a
$\BZ_2$ identification: $$(\phi_1,\phi_2,\phi_3,\phi_4) \sim
(-\phi_1,-\phi_2,-\phi_3,-\phi_4)$$ and thus one has a $\BZ_2$
orbifold singularity at the origin. One can however verify that the
singularity is not a point in the compact Calabi-Yau manifold and
hence the hypersurface does not inherit the singularity  from the
ambient projective space. In a similar fashion, $\BP^{1,1,1,1,4}$
has an $\BZ_4$ orbifold singularity in the origin of the chart
$\phi_5=1$ and $\BP^{1,1,1,2,5}$ has two orbifold singularities:
a $\BZ_5$ in the chart $\phi_5=1$ and a $\BZ_2$ in the chart
$\phi_4=1$. In all these cases, the hypersurface does not inherit the
singularities and hence one obtains a smooth Calabi-Yau manifold.

There might be other situations where the hypersurface does inherit
the singularity and one has the option of blowing up the singularity
by, say a $\BP^1$.
This naturally leads us to models with two K\"ahler moduli where the second
modulus is associated with ``size'' of the blown up $\BP^1$.

\subsection{Two parameter examples}

The GLSM involves seven chiral superfields $\Phi_i$ 
and two abelian vector superfields. We shall again refer to $\Phi_0$
as the $P$-field. There are two K\"ahler moduli which are
related to the Fayet-Iliopoulos terms which we shall label
$t_a = r_a +i {\theta_a\over{2\pi}}$, for $a=1,2$. 
The D-term constraints (moment maps) are
$$
D^a = -e^2 \left(\sum_{i=0}^{5} q_i |\phi_i|^2- r_a\right)\quad,
$$
We shall consider two models (these have been studied in
\cite{twopar}): 
\begin{enumerate}
\item The charge vectors are
$$
q_i^a=\pmatrix{-4&0&0&1&1&1&~1 \cr
            ~0&1&1&0&0&0&-2 }\quad,
$$
and the model is the blowup of the weighted projective space
$\BP^{1,1,2,2,2}$. The compact Calabi-Yau is a
hypersurface of bidegree $(4,0)$ such as
$$
\BP^{1,1,2,2,2}[8]:\quad G=\phi_1^8\phi_6^4 + \phi_2^8\phi_6^4 
+ \phi_3^4 +\phi_4^4 +\phi_5^4 =0
$$
\item The charge vectors are
$$
q_i^a=\pmatrix{-6&0&0&1&1&3&~1 \cr
            ~0&1&1&0&0&0&-2 }\quad,
$$
and the model is the blowup of the weighted projective space
$\BP^{1,1,2,2,6}$. However, it has a $\BZ_6$ singularity 
which is not inherited by the hypersurface given by a transverse
polynomial of bidegree $(6,0)$ such as
$$
\BP^{1,1,2,2,6}[12]:\quad G=\phi_1^{12}\phi_6^6 + \phi_2^{12}\phi_6^6 
+ \phi_3^6 +\phi_4^6 +\phi_5^6 =0
$$
\end{enumerate}

\subsection{Geometric data}

Let $J_a$ ($a=1,\ldots,h_{1,1}$) be the K\"ahler classes  of a
Calabi-Yau manifold with the complexified K\"ahler cone given by
$K=\sum_a t_a J_a$.
The data associated with the special K\"ahler geometry associated with 
$(2,2)$ 
Calabi-Yau compactifications is encoded in a prepotential ${\cal F}(t_a)$
which in the large-volume limit takes the general form\cite{spgeom}
\begin{equation}
{\cal F} = -{1\over6} \kappa_{abc} t_a t_b t_c +{1\over2}\alpha_{ab} t_a t_b +
\beta_{a} t_a + {1\over2}\gamma 
+ {\rm instanton\ corrections}
\end{equation}
where $\alpha_{ab}$, $\beta_a$ and $\gamma$ are constants which reflect
an Sp$(2h_{1,1}+2)$ ambiguity and $\kappa_{abc}$ is the triple
intersection
\begin{equation}
\kappa_{abc} =\int_M J_a \wedge J_b \wedge J_c \equiv \langle J_a J_b
J_c \rangle_M\quad.
\end{equation}
The associated period vector is given by 
\begin{equation}
\vec{\Pi}(t)= \left(\begin{array}{c} 
 {1\over6} \kappa_{abc} t_a t_b t_c + \beta_{a} t_a + \gamma \\[3pt]
-{1\over2} \kappa_{abc} t_b t_c + \alpha_{ab} t_b + \beta_a \\[3pt]
t_a \\[3pt]
1
\end{array}
\right)
\end{equation}
where we have not included the instanton contributions which are
suppressed in the large volume limit.

The central charge associated with a D-brane with RR charge vector
$\vec{n}=(n_6,n_4^a,n_2^a,n_0)$ is given by\footnote{In the one
parameter models, this choice differs in the choice of the sign
of $n_6$ from those considered in \cite{quintic,scheid}. We account
for this when we compare our results with theirs.  We make
this choice in order to have a uniform convention across all models.}
\begin{equation}
Z(\vec{n})= \vec{n}\cdot \vec{\Pi} 
\label{bcharge}
\end{equation}
The central charge associated with a coherent sheaf $E$ is given
by\cite{bpsalgebra}
\begin{equation}
Z(E) = \int_M \ e^{-t_a J_a}\wedge ch(E)\wedge 
\left(1 + {{c_2(M)}\over{24}} \right)
\label{ccharge}
\end{equation}
By comparing the expressions (\ref{bcharge}) and (\ref{ccharge}) one
obtains a map relating the Chern classes of $E$ to the D-brane charges
$\vec{n}$. We obtain
\begin{eqnarray}
{\rm ch}_0(E)&=& - n_6 \nonumber \\
{\rm ch}_1(E)&=& - n_4^a J_a  \\
\langle {\rm ch}_2(E) J_b \rangle_M &=& -n_2^b - n_4^a \alpha_{ab} +
n_6\left(\beta_b - {{\langle c_2(M)J_b\rangle_M}\over{24}}\right) \nonumber \\
\langle {\rm ch}_3(E) \rangle_M  &=& n_0 + \left(\beta_b + {{\langle
c_2(M)J_b\rangle_M}\over{24}}\right) n_4^b + \gamma n_6  \nonumber
\end{eqnarray}
We will now impose the condition that the line bundle ${\cal O}$ correspond
to the pure anti-six brane i.e., $n_6=-1$ and all other $n$'s being zero. This 
is satisfied by the choice
\begin{eqnarray}
\beta_b &=& {{\langle c_2(M)J_b\rangle_M}\over{24}} \nonumber\\
\gamma &=&0
\end{eqnarray}
This does not fix the parameter $\alpha_{ab}$. We will make model-dependent
choices in order to make the two-brane charges $n_2^a$ integers. This
is similar to the requirement that the periods change by an integer
symplectic matrix under the action $t_a\rightarrow t_a+1$ for all
$a$.

\subsubsection{One parameter models}

The intersection numbers for the one parameter models are\cite{kt,font}:
\begin{equation}
\kappa = \langle J^3 \rangle_M = |q_0|\ \langle J^4
\rangle_X
= {{|q_0|}\over{\prod_{i=1}^{5} q_i }}
\end{equation}
Thus we have 
\begin{eqnarray}
\BP^{1,1,1,1,1}[5]:&& \quad \kappa=5,\ \alpha=-{{11}\over2}, \ 
\beta={{25}\over{12}} \\
\BP^{1,1,1,1,2}[6]:&& \quad \kappa=3,\ \alpha=-{9\over2}, \ 
\beta={{7}\over{4}} \\
\BP^{1,1,1,1,4}[8]:&& \quad \kappa=2,\ \alpha=-3, \ \beta={{11}\over{6}} \\
\BP^{1,1,1,2,5}[10]:&& \quad \kappa=1,\ \alpha=-{1\over2}, \ 
\beta={{17}\over{12}} 
\end{eqnarray}
In the above, we have chosen values for $\alpha$ as in ref. \cite{kt} where
the choice enabled the large volume periods of $M$ to be related to the
periods (in the large complex structure limit)
on the mirror $W$ by means of an integer symplectic matrix. In our case,
this seems to ensure that the RR-charges are integers.

\subsubsection{Two parameter examples}

The K\"ahler classes are generated by $H\equiv J_1$ and $L\equiv J_2$. 
The non-vanishing intersection numbers are
\begin{eqnarray}
\kappa_{111} \equiv \langle H^3 \rangle_M = 2|q^1_0|\ \langle H^4 \rangle_X  \\
\kappa_{112} \equiv \langle H^2L \rangle_M = 2|q^1_0|\ \langle H^3 L \rangle_X 
\end{eqnarray}
One also has the relation $L^2=0$ for both models. The data for the two
models are
\begin{eqnarray}
\BP^{1,1,2,2,2}[8]:&& \quad \kappa_{111}=2\kappa_{112}=8,\
\beta_1={7\over3},\ \beta_2=1 \\
\BP^{1,1,2,2,6}[12]:&&\quad\kappa_{111}=2\kappa_{112}=4,\ \beta_1={{13}\over6},
\ \beta_2=1 
\end{eqnarray}
We will choose to set $\alpha_{ab}=0$ in both examples.

\section{Helices and Mutations}

A coherent sheaf $E$ on a  variety $X$ (of dimension $n$)
is called {\em exceptional} if
\begin{eqnarray*}
{\rm Ext}^i(E,E)&=&0\quad,\quad i\geq1 \\
{\rm Ext}^0(E,E)&=&\BC\quad,
\end{eqnarray*}
where Ext${}^i(E,F)$ is the sheaf-theoretic generalisation of the cohomology
groups $H^i(X,E^\ast\otimes F)$ for vector bundles $E$ and $F$.
An ordered collection of exceptional sheaves ${\cal E}=(E_1,\ldots,E_k)$
is called a {\em strongly exceptional collection}\footnote{This is
somewhat different from the definition of Rudakov and is the one
used by Bondal\cite{bondal}. An exceptional collection is one for
which the Ext${}^i(E_a,E_b)=0$ for $a>b$.}
if for all $a<b$, one has
\begin{eqnarray*}
{\rm Ext}^i(E_b,E_a) &=&0 \quad,\ i\geq0  \\
{\rm Ext}^i(E_a,E_b) &=&0 \quad,\ i\neq i_0\quad,
\end{eqnarray*}
for some $i_0$ (which is typically zero).
The alternating sum of dimensions of the groups Ext${}^i(E,F)$ defines a
bilinear product (we call this the {\em Euler form})
\begin{equation}
\chi(E,F)=\sum_{i=0}^{n} (-)^i {\rm dim}\ \left({\rm Ext}^i(E,F) \right)
=\int_X {\rm ch}(E^\ast\otimes F){\rm Td}(X)
\end{equation}
For the case of vector bundles $E$ and $F$,
this is the Witten index associated with the Dolbeault operator on the bundle
$E^\ast\otimes F$ and Td$(X)$ is the Todd class of the tangent bundle of
$X$.

For an exceptional collection, the matrix (``the Euler matrix'')
$$
I_{ab}\equiv \chi(E_a,E_b) \quad,
$$
is an upper-triangular matrix with ones on the diagonal.
For a strongly exceptional collection, the non-zero
entries are given by $(-)^{i_0} {\rm dim}\ {\rm
Ext}^{i_0}(E_a,E_b)$.

New exceptional collections can be generated from old ones by a process
called {\em mutation}. A {\em right mutation} of an exceptional pair
$(E_a,E_{a+1})$ in an exceptional collection is defined by
\begin{equation}
R_{a+1}(E_a,E_{a+1}) = (E_{a+1},R_{E_{a+1}}(E_a))
\end{equation}
and a {\em left mutation} by
\begin{equation}
L_{a}(E_a,E_{a+1}) = (L_{E_{a}}(E_{a+1}),E_a)
\end{equation}
where we have introduced two new sheaves $R_{E_{a+1}}(E_a)$ and
$L_{E_{a}}(E_{a+1})$ which are defined through exact sequences (see 
\cite{rudakov} for details). For example, when Ext${}^0(E_a,E_{a+1})\neq0$
and the (evaluation) map Ext${}^0(E_a,E_{a+1})\otimes E_a \rightarrow E_{a+1}$
is surjective, then $L_{E_{a}}(E_{a+1})$ is defined by
\begin{equation}
0\rightarrow L_{E_{a}}(E_{a+1})\rightarrow {\rm Ext}^0(E_a,E_{a+1})\otimes E_a
\rightarrow E_{a+1} \rightarrow 0
\end{equation}
and when the map Ext${}^0(E_a,E_{a+1})\otimes E_a \rightarrow E_{a+1}$
is injective, one uses
\begin{equation}
0 \rightarrow {\rm Ext}^0(E_a,E_{a+1})\otimes E_a \rightarrow E_{a+1} 
\rightarrow L_{E_{a}}(E_{a+1})
\rightarrow 0
\end{equation}
Similarly, $R_{E_{a+1}}(E_a)$ is defined by (in the surjective case)
\begin{equation}
0\rightarrow E_a \rightarrow {\rm Ext}^0(E_a,E_{a+1})^*\otimes E_{a+1}
\rightarrow R_{E_{a+1}}(E_a)\rightarrow 0
\end{equation}
The Chern characters of the new sheaves are given by
\begin{eqnarray}
\pm{\rm ch}(L_{E_{a}}(E_{a+1}) = {\rm ch}(E_{a+1}) - \chi(E_a,E_{a+1}) {\rm
ch}(E_a) \nonumber \\ 
\pm{\rm ch} (R_{E_{a+1}}(E_a)) = {\rm ch}(E_{a}) - \chi(E_a,E_{a+1}) {\rm
ch}(E_{a+1})
\end{eqnarray}
with the plus sign used for the injective case and the minus sign for the
surjective case. Further, the collection is assumed to be strongly
exceptional (with $i_0=0$) and hence $\chi(E_a,E_{a+1})={\rm
dim~Ext}^0(E_a,E_{a+1})$.

The mutation of a strongly exceptional collection may not continue to be
strongly exceptional.  If the mutated collection is also strongly
exceptional, the mutation is called {\em admissible}. The mutations that
we consider in this paper (in order to generate $S_i$)
are assumed to be admissible though we do not always verify this explicitly.

An exceptional collection $(E_i, \ i\in \BZ)$ is called a {\em helix
of period $p$} if for all $s$ the following condition is satisfied:

All pairs $(E_{s-1},E_s)$,
$(E_{s-2},L^1(E_s)),\ldots,(E_{s-p+1},L^{p-2}(E_s))$ admit left
mutations and $L^{p-1}(E_s))=E_{s-p}$.

Thus a sequence of $(p-1)$ left mutations of a helix brings one back
to an element of the helix modulo a shift of $p$. Each collection
$(E_i,E_{i+1},\ldots,E_{i+p})$ is called a {\em foundation} of the
helix $\{E_i\}$. Any helix is determined uniquely by any of its
foundations. One can also define the helix using right mutations.
We shall henceforth use the term helix for the foundation of a helix
since we will have no need to distinguish them.

\subsection{An example}

An example of a helix on $\BP^n$ is furnished by 
$$
{\cal R}=({\cal O}, \cdots, {\cal O}(n))
$$ 
The Euler matrix is given by
$$
I_{ab}=\left({}_{~b-a}^{n+b-a}\right)
$$

By a sequence of left mutations of ${\cal R}^\ast$
one obtains a {\em mutated} helix
\begin{eqnarray*}
{\cal S}&=&(L^n({\cal O}(n)), L^{n-1}{\cal O}(n-1),\cdots, {\cal O}) \\
&=& (\Omega^n(n),\Omega^{n-1}(n-1),\cdots, {\cal O})
\end{eqnarray*}
where $$L^p({\cal O}(p))\equiv L_{\cal O}L_{{\cal O}(1)}\cdots 
L_{{\cal O}(p-1)}({\cal O}(p))\quad.
$$ 
The bundles which make up the helix
${\cal S}$ can be seen to be the ones
which form the $\sum_al_a=0$ orbit in the $\BC^{n+1}/\BZ_{n+1}$ 
orbifold\footnote{
This statement is somewhat imprecise. The $\sum_al_a=0$ bundles for $\BP^2$
are $({\cal O},-\Omega^1(1),{\cal O}(-1))$ where the minus sign reflects
the K-theory class. This can be corrected by introducing a $(-)^n$ factor
with the left mutation.}.
The left mutation can thus be identified with the generator $g$ of the
quantum $\BZ_{n+1}$ symmetry at the orbifold point.

The dual of the Euler sequence (and related sequences) given by
eqn. (\ref{dualeuler}) can be rewritten  in a form which
suggests generalisations to more non-trivial situations:
\begin{equation}
0\rightarrow L^{p}({\cal O}(p))
\rightarrow {\cal O}^{\oplus ({}^{n+1}_{~p})}\rightarrow
L^{p-1}({\cal O}(p))\rightarrow 0
\end{equation}
where $({}^{n+1}_{~p})$ is to be identified with the Euler form
$\chi({\cal O},L^{p-1}({\cal O}(p-1))\otimes {\cal O}(1))$.

\subsection{Two conjectures}

In the following, we will assume that the Calabi-Yau threefold $M$ 
is given by the transverse
intersection of hypersurfaces in some weighted projective space $X$.
We also assume that $M$ is smooth in the sense that all singularities
that it inherits from $X$ are resolved. 

\noindent{\bf Conjecture 1}
{\em The large-volume monodromy action on ${\cal O}$ in the
ambient variety $X$ produces an
exceptional collection which is the foundation of a helix with 
appropriate period $p$}
\begin{equation}
{\cal R}=\left(R_1={\cal O},R_2,\ldots,R_p\right)\quad.
\end{equation}

As is well known, the large-volume monodromy is typically
given by tensoring of
line bundles which do not change either the (large-volume)
stability as well dimension of moduli space of vector bundles. The
simplest exceptional bundle is the brane which wraps the full space
i.e, ${\cal O}$. Thus, the large-volume monodromy action will
generate a set of exceptional line-bundles. 
The non-trivial part of
the conjecture is that this set leads to a helix. 

We are motivated to consider the structure of helices and their
mutations by the work of Hori, Iqbal and Vafa\cite{HIV} and
earlier work of Zaslow\cite{zaslow}. In this work, they show that
mutations of exceptional collections have a well-defined physical
interpretation in terms of brane creation in the mirror to $X$.
Further, on the mirror side, the large complex structure limit (which is  
identified with the large volume limit under the mirror map) is 
the point (in the 
moduli space of complex structures) with {\em maximal unipotent
monodromy}\cite{lty}. The two-parameter examples studied in
\cite{twopar} illustrate this issue.
The helix structure that we see reflect this structure.
The period of the helix is related to the quantum symmetry
which appears as a classical symmetry on the mirror.

\noindent{\bf Conjecture 2}
{\em All exceptional bundles/sheaves can be obtained by 
mutations of the helix ${\cal R}$ given by Conjecture 1.}
In particular, there exists a
mutated helix ${\cal S}=(S_p,\ldots,S_1={\cal O})$ with
$S_i=L^{i-1}(R_i)$, where $L^{i-1}$ corresponds to a sequence of $(i-1)$
left-mutations.

\noindent {\bf Conjecture 2a}:  {\em All the exceptional bundles on the CY
which correspond to the the $\sum_a l_a =0$ states at the Gepner
point are obtained by the restriction $S_i|_M\equiv V_i$ to the CY
hypersurface.}

The motivation for this conjecture comes from the simple observation,
that $V_i$ is indeed the $\sum_al_a=0$ orbit for the case of 
the quintic in $\BP^4$. Further, it is a suggestive structure which
works in other examples as well (as we will see in the next section). 
Another important motivation comes from the work of Douglas and Diaconescu
which suggests a dual relationship between the  bundles $R_i$
and the bundles $S_i$ i.e., $\chi(R_i,S_j)=\delta_{ij}$. The
process of mutations can be seen to change the upper-triangular
matrix $\chi({\cal R},{\cal R})$ into a diagonal one by mutating
the helix which is the second argument of $\chi({\cal R},{\cal R})$
(as shown in appendix A).
One obvious caveat is that on
restriction to the CY hypersurface, new moduli may appear. 
While one expects no new moduli to appear to the restriction of
${\cal O}$, mirror symmetry predicts the quantum $\BZ_p$ symmetry
which relate the other entries in the helix $S$ and hence we
expect the result to go through for these cases.

\noindent{\bf Conjecture 2 (Stronger form)}
{\em  The helix ${\cal R}$ generates the bounded derived category of
coherent sheaves on $X$, ${\cal D}^b({\rm Coh}(X))$.}

Conjecture 2 (in both forms) is true for $\BP^n$ and seems to follow
from the results of Bondal (see Theorem 4.1 in \cite{bondal})
for spaces with very ample anticanonical class. This implies that
there exists a generalisation of Beilinson's theorem\cite{beilinpn}
for the case of weighted projective spaces along the lines followed by
\cite{gorrud,drezet} for $\BP^n$. The existence of such a generalisation
is also implicit in the work of Douglas and Diaconescu\cite{DD}.

\section{Testing the conjectures}

We first discuss the case of  $\BP^{1,1,1,1,2}[6]$ where we explicitly
show how one deals with the orbifold singularity in the ambient weighted
projective space.  This also illustrates the methods that we employ
in all other examples.

\subsection{$\BP^{1,1,1,1,2}[6]$}

The Todd class for $X=\BP^{q_1,q_2,q_3,q_4,q_5}$ is given by
$$
{\rm Td}(X) = \prod_{i=1}^5 \left({q_iJ\over{1-e^{-q_iJ}}}\right)
$$

For the one parameter models, we expand the Chern character of a sheaf
$E$ as
$$
{\rm ch}(E)=Q_0 + Q_1 J + Q_2 {{J^2}\over2} + Q_3 {{J^3}\over6}\quad.
$$
The translation of the $Q_a$ to the RR charges is
\begin{eqnarray}
n_6 &=& - Q_0 \nonumber \\
n_4 &=& - Q_1 \\
n_2 &=& - {\kappa\over2}Q_2 + \alpha Q_1 \nonumber \\
n_0 &=& {\kappa\over6}Q_3 + 2 \beta Q_1 \nonumber 
\end{eqnarray}
where $\kappa$, $\alpha$ and $\beta$ are as defined earlier.

The large volume monodromy is obtained by
the operations $\theta \rightarrow \theta + 2\pi$ in the GLSM. 
For the one parameter models, one
obtains ${\cal O}\rightarrow {\cal O}(J)$.
Thus, conjecture one predicts that the line bundles ${\cal O}(bJ)$
should form a helix.  However, as pointed out earlier, the ambient
weighted projective space has a $\BZ_2$ singularity which is
not inherited by the Calabi-Yau hypersurface. The {\em naive} computation
of Euler form gives rise to fractions. The occurrence of these
fractions is related to orbifold singularity. We are interested in
getting rid of the contribution from the singularity. We modify
the Euler form as follows\footnote{We do not explicitly calculate
this contribution since it will take us too far from our objective. It
suffices to note that one observes a $\BZ_2$ pattern in the unmodified
numbers associated with the Euler form
and this simple modification leads to an upper
triangular matrix with ones on the diagonal. The true justification for
this is that we get sensible results on restricting the bundles to the
Calabi-Yau hypersurface.}:
\begin{equation}
\widehat{\chi}\left({\cal O},{\cal O}(bJ)\right) \equiv 
\chi\left({\cal O},{\cal O}(bJ)\right)
+{{(-)^b}\over{32}}
\end{equation}
By explicitly using the {\em modified} Euler form,
we obtain the following helix ${\cal R}$ of period six given by
$$
{\cal R} = \left[{\cal O}, {\cal O}(J), {\cal O}(2J), {\cal O}(3J),
{\cal O}(4J), {\cal O}(5J)\right]
$$
The period of length six reflects the quantum $\BZ_6$ symmetry of the
model.
The Euler matrix is 
\begin{equation}
I_{ab}=\widehat{\chi}(R_a,R_b)=
\pmatrix{1&4&11&24&46&80 \cr
0&1&4&11&24&46\cr
0&0&1&4&11&24\cr
0&0&0&1&4&11\cr
0&0&0&0&1&4\cr
0&0&0&0&0&1}
\end{equation}
One can also explicitly verify the numbers by counting the number of
degree one monomial, degree two monomials and so on. For example,
there are four degree one monomials  $(\phi_1,\phi_2,\phi_3,\phi_4)$ 
and eleven degree two monomials (ten quadratic in $\phi_i$
$i=1,\ldots,4$ and $\phi_5$) and so on.

We need to now verify the second conjecture i.e., the $\sum_a l_a=0$ bundles
are given by a sequence of left mutations. The bundle $-S_2=L_{\cal O}({\cal
O}(J))$ is defined by the exact sequence 
\begin{equation}
0\rightarrow L_{\cal O}({\cal O}(J)) \rightarrow  {\cal O}^{\oplus 4}
\stackrel{f}{\rightarrow} {\cal O}(J) \rightarrow 0
\end{equation}
where $f^i=\phi_i$ for $i=1,2,3,4$. This sequence is similar to the
Euler sequence associated with $\BP^3$ (with homogeneous coordinates
$\phi_1,\ldots,\phi_4$) and is a bundle of rank three. 
The next bundle is given by the exact sequence
\begin{equation}
0\rightarrow S_3
\rightarrow {\cal O}^{\oplus 5}\rightarrow
L_{\cal O}({\cal O}(J))\otimes{\cal O}(J)\rightarrow 0
\end{equation}
We now come to the next one in the sequence. This is an
interesting one. One can see that
Hom$\left[{\cal O}, S_3\otimes {\cal O}(J)=L_{{\cal O}(1)} L_{{\cal O}(2)}
{\cal O}(3)\right]$ vanishes. Thus, the map is injective unlike the earlier
ones and we obtain the sequence
\begin{equation}
0\rightarrow S_3\otimes {\cal O}(J) \rightarrow  S_4 \rightarrow 0
\end{equation}
This implies that $S_4=S_3\otimes {\cal O}(J)$.
The other bundles are obtained in similar fashion. 
Without getting into sequences involved, one 
can easily obtain the Chern character of the bundles by
using the inverse of the Euler matrix given above (following
the argument in the appendix)
The inverse matrix is given by
\begin{equation}
I^{-1}_{ab} =\pmatrix{1& -4&  5&  0& -5&  4\cr 
                      0&  1& -4&  5&  0& -5\cr 
                      0&  0&  1& -4&  5&  0\cr
                      0&  0&  0&  1& -4&  5\cr 
                      0&  0&  0&  0&  1& -4\cr 
                      0&  0&  0&  0&  0&  1}
\end{equation}
The Chern character of the bundles are
\begin{eqnarray*}
{\rm ch}(S_2)&=& {\rm ch}({\cal O}(J)) -4 \\
{\rm ch}(S_3)&=& {\rm ch}({\cal O}(2J)) -4{\rm ch}({\cal O}(J)) +5 \\
{\rm ch}(S_4)&=& {\rm ch}({\cal O}(3J)) -4{\rm ch}({\cal O}(2J)) +5 {\rm
ch}({\cal O}(J))\\
{\rm ch}(S_5)&=& {\rm ch}({\cal O}(4J)) -4{\rm ch}({\cal O}(3J)) +5 {\rm
ch}({\cal O}(2J)) -5\\
{\rm ch}(S_6)&=& {\rm ch}({\cal O}(5J)) -4{\rm ch}({\cal O}(4J)) +5 {\rm
ch}({\cal O}(3J)) -5{\rm ch}({\cal O}(J)) +4 
\end{eqnarray*}
Note that $S_6=L^5({\cal O}(5J))={\cal O}(-J)$ which shows that the
helix has period six. Further, there is a version of Serre duality
which takes the form $S_i \simeq S_{p+1-i}^\ast \otimes S_p$, where $p$ 
is the period of the helix.
Let $V_i=S_i|_M$ be the restriction of these bundles to the 
compact Calabi-Yau
manifold $M$ given by degree six hypersurface in the weighted projective
space. A non-trivial test of our conjectures is to verify that the
RR charges associated with these branes are indeed identical to the ones in
the $\sum_a l_a=0$ orbit. The Chern character of the $\sum_a l_a=0$ bundles are predicted
by our conjectures to be
\begin{eqnarray}
 {\rm ch}(V_1)&=&1  \nonumber \\
 {\rm ch}(V_2)&=&-3 + J + {\frac{{J^2}}{2}} + {\frac{{J^3}}{6}}\nonumber  \\
 {\rm ch}(V_3)&=&2 - 2\,J + {\frac{2\,{J^3}}{3}} \nonumber \\
 {\rm ch}(V_4)&=&2 - J^2 \\
 {\rm ch}(V_5)&=&-3 + 2\,J - {\frac{2\,{J^3}}{3}} \nonumber \\
 {\rm ch}(V_6)&=&1 - J + {\frac{{J^2}}{2}} - {\frac{{J^3}}{6}}\nonumber  
\end{eqnarray}
The Chern character suggest the following relationships among the various
bundles that appear: $V_4=V_3^\ast\otimes {\cal O}(-J)$, $V_5=
V_2^\ast \otimes {\cal O}(-J)$ and $V_6 = {\cal O}(-J)$.
The RR charges are (in the convention $\vec{n}=(n_6,n_4,n_2,n_0)$)
\begin{eqnarray}
\vec{n}_1 &=& (-1, 0, 0, 0) \nonumber \\
\vec{n}_2 &=& (3, -1, -6, 4) \nonumber \\
\vec{n}_3 &=& (-2, 2, 9, -5)\nonumber  \\
\vec{n}_4 &=& (-2, 0, 3, 0) \\
\vec{n}_5 &=& (3, -2, -9, 5) \nonumber \\
\vec{n}_6 &=& (-1, 1, 3, -4)\nonumber  
\end{eqnarray}
This is in agreement with the charges for the corresponding Gepner
model\cite{scheid} (modulo a sign difference in the six-brane
charge). The non-trivial part 
of the
check is that the six-charges that we obtain form a $\BZ_6$ orbit
given by
the action of a matrix $A$ (with $A^6=1$)
with the property $\vec{n}_{i+1} = 
\vec{n}_{i}\cdot A$. This result clearly does not depend on the
choice
of $\alpha$ and reflects the quantum $\BZ_6$ symmetry of the model.
However, our choice of $\alpha$ ensures that the $A$ is a matrix with
integer entries.

\subsection{Other one parameter models}

\subsubsection{The Quintic}

The helix in $\BP^4$ has period five (which reflects the quantum
$\BZ_5$ symmetry)
$$
{\cal R} = \left[{\cal O}, {\cal O}(J), {\cal O}(2J), {\cal O}(3J),
{\cal O}(4J)\right]
$$
The mutated helix corresponding to the the $\sum_a l_a=0$ orbit is given by
$$
S_{i+1}=(-)^i L^{i}({\cal O}({iJ}))=
(-)^i \Omega^i(i)
$$
where $\Omega$ is the cotangent bundle to $\BP^4$ and $\Omega^p$ is
the $p-$th exterior power of the cotangent bundle.
One can verify that the definition of a left mutation leads to the
following exact sequences which are derivable from the Euler sequence
\begin{equation}
0\rightarrow \Omega^{p}(p)
\rightarrow {\cal O}^{\oplus ({}^{5}_{p})}\rightarrow
\Omega^{p-1}(p)\rightarrow 0
\label{dualeuler}
\end{equation}
The restriction of the above sequence to the quintic hypersurface gives
the $\sum_a l_a=0$ orbit as has already been observed in ref. \cite{dfr2}.

\subsubsection{$\BP^{1,1,1,1,4}[8]$}

The foundation of a helix of period eight is found to be 
$$
{\cal R} = \left[{\cal O}, {\cal O}(J), {\cal O}(2J), {\cal O}(3J),
\cdots,{\cal O}(6J), {\cal O}(7J)\right]
$$
In the above, we used the modified Euler form
\begin{equation}
\widehat{\chi}\left({\cal O},{\cal O}(bJ)\right) \equiv 
\chi\left({\cal O},{\cal O}(bJ)\right)
+ {\rm orb}_4(b) 
\end{equation}
$$
{\rm orb}_4(b) =\left\{\matrix{
-7/64 & {\rm for} & b=0~~&{\rm mod}~4 \cr
-1/64 & {\rm for} & b=1,3&{\rm mod}~4 \cr
9/64 & {\rm for} & b=2~~&{\rm mod}~4 } \right.
$$
The term reflects the $\BZ_4$ singularity in
the weighted projective space $\BP^{1,1,1,1,4}$. The
Euler form is given by
\begin{equation}
I_{ab}=\pmatrix{1&4&10&20&36&60&94&140 \cr
0&1&4&10&20&36&60&94\cr
0&0&1&4&10&20&36&60\cr
0&0&0&1&4&10&20&36\cr
0&0&0&0&1&4&10&20\cr
0&0&0&0&0&1&4&10\cr
0&0&0&0&0&0&1&4\cr
0&0&0&0&0&0&0&1}
\end{equation}

The Chern character of the $\sum_a l_a=0$ bundles are
\begin{eqnarray}
 {\rm ch}(V_1) &=& 1  \nonumber \\
    {\rm ch}(V_2) &=& -3 + J + {\frac{{J^2}}{2}} +
{\frac{{J^3}}{6}}\nonumber \\
    {\rm ch}(V_3) &=& 3 - 2\,J + {\frac{2\,{J^3}}{3}}\\
    {\rm ch}(V_4) &=& -1 + J - {\frac{{J^2}}{2}} + {\frac{{J^3}}{6}}
\nonumber
\end{eqnarray}
The Chern character of bundles have the pattern ${\rm ch}V_{i+4}=-{\rm
ch}V_i$ and hence relate branes to anti-branes. We thus have given
the Chern character for only the first four bundles. The Chern character
suggest the following relationship among the bundles: $V_3 = -V_2^\ast
\otimes {\cal O}(-J)$ and $V_4=-{\cal O}(-J)$.
The corresponding RR charges are
\begin{eqnarray}
\vec{n}_1 &=& (-1,0,0,0) \nonumber \\
\vec{n}_2 &=& (3, -1, -4, 4) \nonumber \\
\vec{n}_3 &=& (-3, 2, 6, -6) \\
\vec{n}_4 &=& (1, -1, -2, 4)\nonumber  
\end{eqnarray}
Again, these agree with the corresponding Gepner model results
modulo the sign of $n_6$\cite{scheid}.

\subsubsection{$\BP^{1,1,1,2,5}[10]$}

The helix of period ten is found to be 
$$
{\cal R} = \left[{\cal O}, {\cal O}(J), {\cal O}(2J), {\cal O}(3J),
\cdots,{\cal O}(8J), {\cal O}(9J)\right]
$$
after using the modified Euler form
\begin{equation}
\widehat{\chi}\left({\cal O},{\cal O}(bJ)\right) \equiv 
\chi\left({\cal O},{\cal O}(bJ)\right)
+ {\rm orb}_2(b) + {\rm orb}_5(b)
\end{equation}
where ${\rm orb}_2(b)=(-)^b/32$ and 
$$
{\rm orb}_5(b) =\left\{\matrix{
-2/25 & {\rm for} & b=0,1,4&{\rm mod}~5 \cr
3/25 & {\rm for} & b=2,3~~&{\rm mod}~5 }\right.
$$
The two terms reflect the $\BZ_2$ and $\BZ_5$ singularities in
the weighted projective space $\BP^{1,1,1,2,5}$. The modified
Euler matrix is found to be
\begin{equation}
I_{ab}=\pmatrix{1&3&7&13&22&35&53&77&108&147\cr 
0&1&3&7&13&22&35&53&77&108\cr
0&0&1&3&7&13&22&35&53&77\cr
0&0&0&1&3&7&13&22&35&53\cr
0&0&0&0&1&3&7&13&22&35\cr
0&0&0&0&0&1&3&7&13&22\cr
0&0&0&0&0&0&1&3&7&13\cr
0&0&0&0&0&0&0&1&3&7\cr
0&0&0&0&0&0&0&0&1&3\cr
0&0&0&0&0&0&0&0&0&1}
\end{equation}

We now give the exact sequences which define $-S_2$ and $S_3$
\begin{equation}
0\rightarrow (-S_2) \rightarrow  {\cal O}^{\oplus 3}
\stackrel{f}{\rightarrow} {\cal O}(J) \rightarrow 0
\end{equation}
where $f^i=\phi_i$ for $i=1,2,3$. 
The next bundle is given by the exact sequence
\begin{equation}
0\rightarrow S_3
\rightarrow {\cal O}^{\oplus 2}\rightarrow
(-S_2)\otimes{\cal O}(J)\rightarrow 0
\end{equation}

The $\sum_a l_a=0$ bundles are
\begin{eqnarray}
{\rm ch}(V_1) &=& 1 \nonumber \\
{\rm ch}(V_2) &=& -2 + J + {\frac{{J^2}}{2}} + {\frac{{J^3}}{6}}\nonumber  \\
{\rm ch}(V_3) &=& -J + {\frac{{J^2}}{2}} + {\frac{5\,{J^3}}{6}} \nonumber \\
{\rm ch}(V_4) &=& 2 - J - {\frac{{J^2}}{2}} + {\frac{5\,{J^3}}{6}} \\
{\rm ch}(V_5) &=& -1 + J - {\frac{{J^2}}{2}} + \frac{{J^3}}{6}\nonumber  
\end{eqnarray}
Observe that $V_3$ is a sheaf.
The Chern character of sheaves have the pattern ${\rm ch}V_{i+5}=-{\rm
ch}V_i$ and hence relate branes to anti-branes. We thus have given
the Chern character for only the first five sheaves. 
The corresponding RR charges are
\begin{eqnarray}
\vec{n}_1 &=& (-1,0,0,0) \nonumber \\
\vec{n}_2 &=& (2, -1, -1, 3) \nonumber \\
\vec{n}_3 &=& (0, 1, 0, -2) \\
\vec{n}_4 &=& (-2, 1, 1, -2) \nonumber \\
\vec{n}_5 &=& (1, -1, 0, 3)\nonumber  
\end{eqnarray}
Again, these agree with the corresponding Gepner model results
modulo the sign of $n_6$\cite{scheid}.

\subsection{The two parameter models}
The Todd class is defined by
$$
{\rm Td}(X) = \left({H\over{1-e^{-H}}}\right)^2
\left({qH\over{1-e^{-qH}}}\right)
              \left({L\over{1-e^{-L}}}\right)^2
              \left({{H-2L}\over{1-e^{2L-H}}}\right)
$$
where $q=1$ for $\BP^{1,1,2,2,2}$ and $q=3$ for $\BP^{1,1,2,2,6}$.
One has the intersection relations 
\begin{eqnarray*}
X=\BP^{1,1,2,2,2}:\quad&& \langle H^4 \rangle_X = 2 \langle H^3 L
\rangle_X=2,\ L^2=0 \\
X=\BP^{1,1,2,2,6}:\quad&& \langle H^4 \rangle_X = 2 \langle H^3 L
\rangle_X=2/3,\ L^2=0 
\end{eqnarray*}

We define $h$ and $l$ (which generate $H_2(M)$) satisfying
\begin{eqnarray*}
\langle h\cdot H \rangle_M = 1 \qquad \langle h \cdot L \rangle_M =0 \\
\langle l\cdot H \rangle_M = 0 \qquad \langle l \cdot L \rangle_M =1
\end{eqnarray*}
It is easy to work out the relationship between $(h,l)$  and $(H^2,HL)$.
The Chern character of a bundle $E$ is given in terms of the RR-charges
of the associated D-brane by (when $\alpha_{ab}$ are set to zero)
\begin{eqnarray}
{\rm ch}_0(E) &=& -n_6 \nonumber \\
{\rm ch}_1(E) &=& -n_4^1 H - n_4^2 L \nonumber \\
{\rm ch}_2(E) &=&-n_2^1 h - n_2^2 l \nonumber \\
{\rm ch}_3(E) &=& n_0 + 2 \beta_1 n_4^1 +2 \beta_2 n_4^2
\end{eqnarray}

Both models that we consider are K3 fibrations over a base $\BP^1$.
The K3 fibre in the case of  $\BP^{1,1,2,2,2}[8]$ is a quartic in
in $\BP^3$ while the K3 fibre in the case of $\BP^{1,1,2,2,6}[12]$
is a degree six surface in $\BP^{1,1,1,3}$.
We will see that the helix structure reflects this structure. 
The helix will can be obtained (loosely-speaking) from the tensor
product of two helices
$$
{\cal R} = \left({\cal O}, {\cal O}(L)\right) \otimes 
\left({\cal O},{\cal O}(H),\cdots,{\cal O}(d H)\right)
$$
where the first helix is associated with the base and the other with
the K3 fibre.

\subsubsection{$\BP^{1,1,2,2,2}[8]$}

Under the operations $\theta_i \rightarrow \theta_i + 2\pi$, one
obtains ${\cal O}\rightarrow {\cal O}(H)$ and ${\cal O}\rightarrow{\cal O}(L)$.
Thus, conjecture one predicts that the line bundles ${\cal O}(aH+bL)$
should form a helix. By explicitly computing the Euler form,
we obtain the following helix of period eight, ${\cal R}$ given by
$$
{\cal R} = \left[{\cal O}, {\cal O}(L), {\cal O}(H), {\cal O}(H+L),
{\cal O}(2H), {\cal O}(2H+L), {\cal O}(3H), {\cal O}(3H+L) \right]
$$
with its Euler matrix being upper triangular as required.
\begin{equation}
I_{ab}=\chi(R_a,R_b)=
\pmatrix{1&2&6&10&20&30&50&70 \cr
0&1&2&6&10&20&30&50\cr
0&0&1&2&6&10&20&30\cr
0&0&0&1&2&6&10&20\cr
0&0&0&0&1&2&6&10\cr
0&0&0&0&0&1&2&6\cr
0&0&0&0&0&0&1&2\cr
0&0&0&0&0&0&0&1}
\end{equation}

We need to now verify the second conjecture i.e., the $\sum_a l_a=0$ bundles
are given by a sequence of left mutations. The bundle $L_{\cal O}({\cal
O}(L))$ is defined by the exact sequence 
\begin{equation}
0\rightarrow L_{\cal O}({\cal O}(L)) \rightarrow  {\cal O}^{\oplus 2}
\stackrel{J}{\rightarrow} {\cal O}(L) \rightarrow 0
\end{equation}
where $J^i=\phi_i$ for $i=1,2$. The line bundle obtained this
way can be seen to equal ${\cal O}(-L)$. 

The Chern character of the bundles obtained on restriction to
the Calabi-Yau hypersurface are given by
\begin{eqnarray}
 {\rm ch}(V_1)&=&1 \nonumber \\
{\rm ch}(V_2)&=&-1 + L \nonumber \\
    {\rm ch}(V_3)&=&-3  + H - 2L+ 4h+ 2l  +{4\over3} \nonumber \\
    {\rm ch}(V_4)&=&3 - H - L- 2l  +{2\over3}\\
    {\rm ch}(V_5)&=&3   - 2H + 4L - 8h+ {4\over3} \nonumber \\
    {\rm ch}(V_6)&=&-3 + 2H - L -{4\over3}\nonumber \\
    {\rm ch}(V_7)&=&-1 + H -3L+4h- 2l  - {8\over3} \nonumber \\
 {\rm ch}(V_8)&=&1 - H + L+ 2l  +{2\over3} \nonumber 
\end{eqnarray}
These results agree with the Gepner model results\cite{scheid} and 
also agree
with the bundles given by Douglas and Diaconescu\cite{DD}.

\subsubsection{$\BP^{1,1,2,2,6}[12]$}

Under the operations $\theta_i \rightarrow \theta_i + 2\pi$, one
obtains ${\cal O}\rightarrow {\cal O}(H)$ and ${\cal O}\rightarrow{\cal
O}(L)$.
Thus, conjecture one predicts that the line bundles ${\cal O}(aH+bL)$
should form a helix. By explicitly computing the Euler form,
we obtain the following helix of period twelve, ${\cal R}$ given by
$$
{\cal R} = \left[{\cal O}, {\cal O}(L), {\cal O}(H), {\cal O}(H+L),
\cdots, {\cal O}(5H), {\cal O}(5H+L) \right]
$$
The ambient weighted projective space $X$ has as $\BZ_6$ singularity 
and we need to define a modified Euler form.
The modified Euler form is given by
\begin{equation}
\widehat{\chi}\left(R_a,R_b\right) \equiv
\chi\left(R_a,R_b\right)
+ {\rm orb}_6(b-a) 
\end{equation}
where
$$
{\rm orb}_6(b) =\left\{\matrix{
-2/27 & {\rm for} & b=0,1,5&{\rm mod}~6 \cr
1/27 & {\rm for} & b=2,4~~&{\rm mod}~6 \cr
4/27 & {\rm for} & b=3~~~~&{\rm mod}~6 }\right.
$$

\begin{equation}
\widetilde{\chi}(R_a,R_b)=
\pmatrix{
1&2&5&8&14&20&31&42&60&78&105&132 \cr
0&1&2&5&8&14&20&31&42&60&78&105 \cr
0&0&1&2&5&8&14&20&31&42&60&78 \cr
0&0&0&1&2&5&8&14&20&31&42&60 \cr
0&0&0&0&1&2&5&8&14&20&31&42 \cr
0&0&0&0&0&1&2&5&8&14&20&31 \cr
0&0&0&0&0&0&1&2&5&8&14&20 \cr
0&0&0&0&0&0&0&1&2&5&8&14 \cr
0&0&0&0&0&0&0&0&1&2&5&8 \cr
0&0&0&0&0&0&0&0&0&1&2&5 \cr
0&0&0&0&0&0&0&0&0&0&1&2 \cr
0&0&0&0&0&0&0&0&0&0&0&1}
\end{equation}
One can then determine the Chern characters associated with
the restrictions of the $S_i$ to the Calabi-Yau hypersurface.
We obtain
\begin{eqnarray}
   {\rm ch}(V_1) &=& 1  \nonumber \\
 {\rm ch}(V_2) &=& -1 + L  \nonumber \\
   {\rm ch}(V_3) &=& -2  + H-2L+ 2h  + l  +{2\over3} \nonumber \\
   {\rm ch}(V_4) &=& 2  - H - l +{1\over3} \\
   {\rm ch}(V_5) &=& 1  - H + 2L- 2h+ l  +{4\over3}\nonumber \\
   {\rm ch}(V_6) &=& -1 + H - L - l -{1\over3} \nonumber 
\end{eqnarray}
Note that ch$V_i=-$ch$V_{i+6}$ and hence we have not listed the
other six cases. The Chern character suggest the following
identification among various bundles: $V_{i+1}=-V_{i}\otimes {\cal
O}(-L)$ for $i=1,3,5$ and $V_5={\cal O}(2L-H)$.
The RR charges associated with the above vector bundles
are (in the convention $(n_6,n_4^1,n_4^2,n_0,n_2^1,n_2^2)$)
\begin{eqnarray}
 \vec{n}_1&=& (-1, 0, 0, 0, 0, 0) \nonumber \\
 \vec{n}_2&=&(1, 0, -1, 2, 0, 0) \nonumber \\
 \vec{n}_3&=&(2, -1, 2, 1, -2, -1) \nonumber \\
 \vec{n}_4&=&(-2, 1, 0, -4, 0, 1) \\
 \vec{n}_5&=&(-1, 1, -2, 1, 2, -1) \nonumber \\
 \vec{n}_6&=&(1, -1, 1, 2, 0, 1)\nonumber  
\end{eqnarray}
This agrees with the corresponding numbers obtained in \cite{scheid}
for the corresponding Gepner model.

\section{Quivers from Helices and onward}
\subsection{The Beilinson quiver}

Associated with the foundation of a helix ${\cal R}$,
made up of a strongly exceptional collection, one can obtain a quiver
in the following manner. If ${\cal R}= \oplus R_i$, and defining 
$A = {\rm Hom} ({\cal R}, {\cal R})$, we can identify $A$ as the algebra of 
a quiver with relations. The corresponding quiver diagram may be identified
as follows. The $i$-th vertex is associated with $R_i$, or more precisely
with Hom $(R_i, R_i)$. Draw dim(Hom$(R_i,R_j)$) arrows beginning at
vertex $i$ and ending at vertex $j$.  The quiver
relations are the obvious ones, namely those given by considering
explicitly the maps involved in Hom$(R_i,R_j)$. In the case of $\BP^n$,
where the $R_i={\cal O}(i-1)$, the relations can be obtained by
the obvious rule that Hom$({\cal O}(i), {\cal O}(i + 1))$ is given by 
multiplication by the coordinates of $\BP^n$. Thus for homomorphisms
from ${\cal O}$ to ${\cal O}(2)$, we can have maps that are either of
the form $\phi_1\phi_2$ or equivalently $\phi_2 \phi_1$ and 
so on with the other
coordinates. Thus if links are labelled $X^a_{i,i+1}$, where
$a=1,\ldots (n+1)$
are related to the coordinates and $i=1, \ldots (n+1)$ labelled the
appropriate node in the quiver, we get the relations
\begin{equation}
X^a_{i,i+1}X^b_{i+1,i+2}= X^b_{i,i+1}X^a_{i+1,i+2}
\end{equation}
Note that there is no link between the $(n+1)$-th node of the quiver and the
first node. We refer to this kind of quiver as a generalized 
Beilinson quiver
following the terminology used for the $\BP^2$ case by 
Douglas et. al. in ref. \cite{dfr2}.

The structure of the quiver and its relations become more complicated
for the cases of the weighted projective cases and with extra K\"ahler
moduli. However the rules for working them out remain the same.
Even for the one K\"ahler modulus cases we may have links that
connect to other than next-nearest-neighbour vertices. Such links are
obviously due to maps using the coordinates of weighted projective
space that have higher weight. 

As an illustration we give below two quivers. The first one is for the
case of $\BP^2$ (figure 1)  and the second for the case of $\BP^{1,1,1,1,2}$
(figure 2).
\begin{figure}[ht]
\begin{center}
\leavevmode\epsfysize=2.5cm \epsfbox{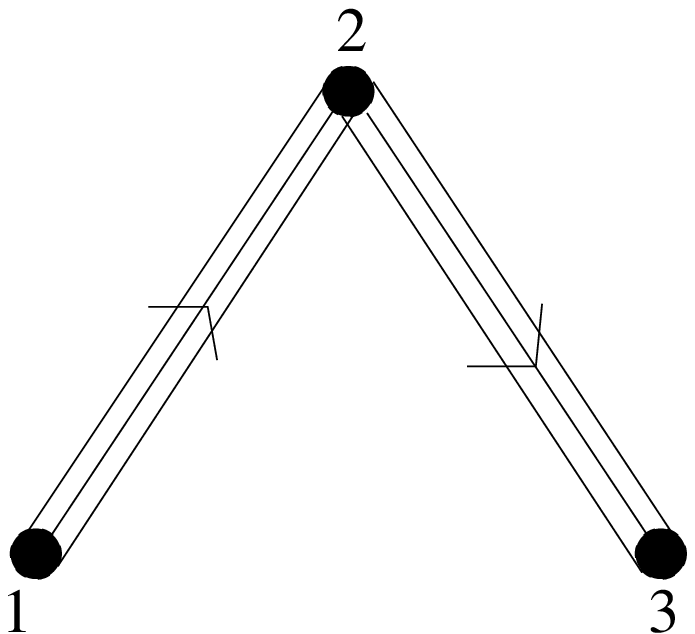}
\end{center}
\caption{Beilinson quiver for $\BP^2$}
\begin{center}
\leavevmode\epsfxsize=4cm \epsfbox{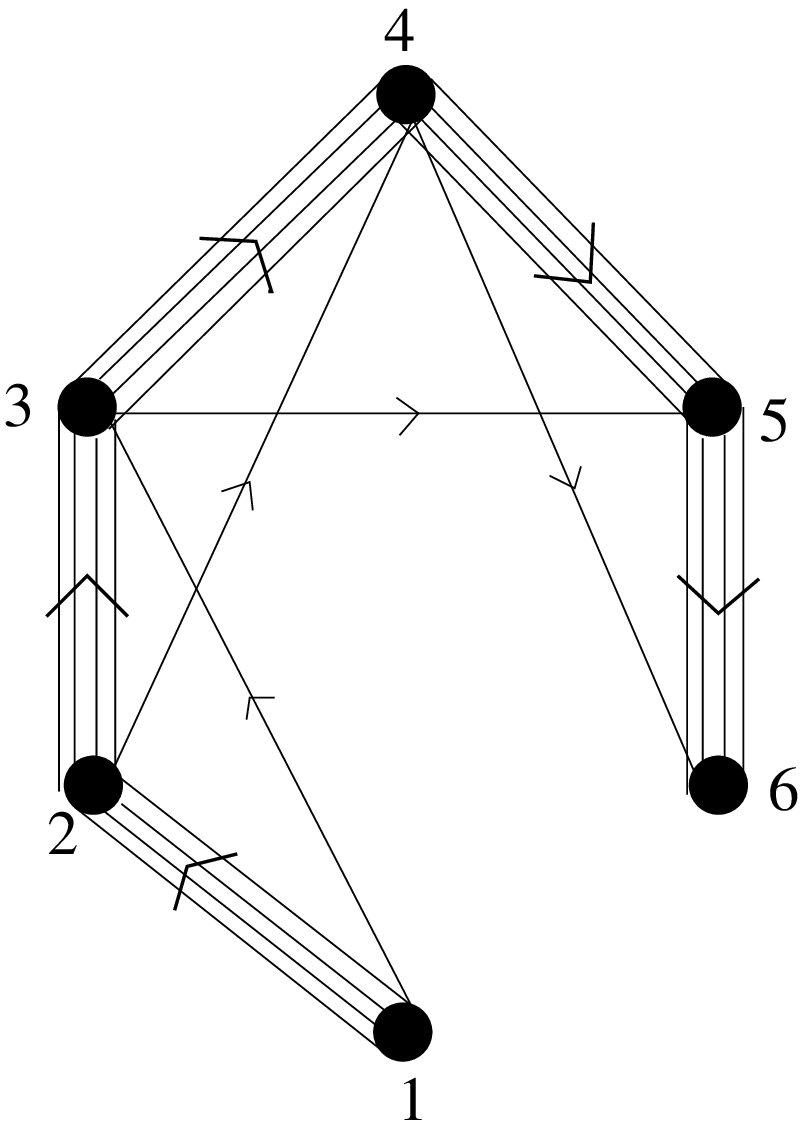}
\end{center}
\caption{Beilinson quiver for $\BP^{1,1,1,1,2}$}
\end{figure}

A representation of a Beilinson quiver for $\BP^n$ is characterised by a
dimension vector $(u_1, u_2, \ldots, u_{n+1})$ which can be associated to a
complex, the Beilinson complex 
\begin{eqnarray}
\lefteqn{0 \rightarrow \BC^{u_1}\otimes \Omega^n(n) 
\stackrel{{X}_{1,2}}{\longrightarrow} 
\BC^{u_2}\otimes \Omega^{n-1}(n-1) \stackrel{X_{2,3}}{\longrightarrow} 
\BC^{u_3}\otimes \Omega^{n-2}(n-2)  \rightarrow \cdots} \hspace{1.6in}
\nonumber \\
&&\cdots \rightarrow \BC^{u_n}\otimes \Omega^1(1) \rightarrow 
\BC^{u_{n+1}}\otimes {\cal O} \rightarrow 0 \quad.
\label{beilcomp}
\end{eqnarray}
where $V=\BC^{n+1}$ with basis $(e_a)$. 
The maps $X_{i,i+1}$ are given by $X^{a}_{i,i+1}e_a$ 
This turns the quiver relations into the statement
$X_{i,i+1}.X_{i+1,i+2}=0$ as is to be expected if the sequence above is
to be a complex. 
If the sequence is exact (for instance) at all nodes except 
one, we
can write the Chern character for the sheaf ${\cal E}$ corresponding to
the appropriate quiver representation as  
\begin{equation}
\label{beilchern}
\pm {\rm ch}({\cal E})= \sum_i (-1)^iu_i\ {\rm ch}(\Omega^{n-i}(n-i)) 
\end{equation}
Thus we may write characterise any sheaf ${\cal E}$ that arises as the
cohomology of a complex made up of the $\Omega^i(i)$, by a vector 
$(u_1,\ldots,u_{n+1})$. The charges of the corresponding D-brane 
configuration
may be read off using the formula (\ref{beilchern}).

The above complex does not generate all sheaves on $\BP^n$. The more
general situation has been considered by Beilinson\cite{beilinpn}.
He has shown that any sheaf ${\cal E}$ on $\BP^n$ is
obtained from a complex $K^{\bullet}$ with $H^0(K^\bullet)={\cal E}$, and
$H^i(K^\bullet)=0$ for $i\neq0$. The $i$-th term of the complex
is isomorphic to $\oplus_j H^{i+j}({\cal E}(-j))\otimes \Omega^j(j)$.
Thus, the Chern character of a generic sheaf (in the large volume limit) 
is again given by the formula (\ref{beilchern}) with an important
difference in that the $u_i$ can be both positive and negative. 
However from the point of view
of the corresponding quiver gauge theory it is clear that we are
restricted to only positive values for {\em all} $u_i$ or negative values 
for all $u_i$. 

We may suggestively re-write both the complex (\ref{beilcomp}) and the 
corresponding formula for an arbitrary sheaf ${\cal E}$ 
by replacing every $\Omega^i(i)$ by $L^i({\cal O}(i))$ in the notation of
Sec. 3 of this paper. Similarly, the $i-$th term in the Beilinson complex
can be rewritten as $\oplus_j H^{i+j}(R_j^\ast\otimes {\cal E})
\otimes S_j$.
This is the form in which we expect that 
the Beilinson complex and Beilinson's theorem to generalize for the more
complicated  cases under consideration in this paper. We note here 
that the  re-writing of Beilinson's theorem in this 
language was pointed out by Gorodontsev and Rudakov \cite{gorrud}
and also appears for instance in the work of Dr\'ezet \cite{drezet}.

The fact that such a generalization is possible is suggested strongly by
the work of Bondal \cite{bondal} \cite{bondal1}. In \cite{bondal} Bondal
shows that  bounded derived category of coherent sheaves on a variety
$X$, denoted ${\cal D}^b({\rm Coh}(X))$ is equivalent to the bounded derived 
category of right modules of an algebra $A$, denoted by ${\cal
D}^b({\rm mod}-A)$. In fact, in our construction of the generalized
Beilinson quiver from the helix, 
we have already implicitly assumes this categorical equivalence.
From Bondal's work, it is clear that the $\{S_i\}$ (called left-dual
by him)  that we have
defined from the $\{R_i\}$ are in fact related to 
the irreducible representations of the quiver algebra. Note that in
formula (\ref{beilchern}) we have explicitly included a sign in the
formula for ${\rm ch}({\cal E})$ on the right hand side. 
If we had considered the objects $S_i$ rather
than the $\Omega^i(i)$, all the signs would have been fixed to be positive.

We can thus proceed to state the result for the characterisation of all
sheaves ${\cal E}$ that arises as the cohomology of a complex of sheaves
whose elements are the $S_i$ that we have written down in the general
case. All such sheaves
are characterised by a vector $(u_1,\ldots,u_n)$, where $n$ is fixed
suitably by the actual variety under consideration. Further
\begin{equation}
\label{genbeilchern}
{\rm ch}({\cal E})= \sum_i u_i\ {\rm ch}(S_i)
\end{equation}
In the large-volume limit, the numbers $u_i$ can be positive or
negative, whereas at the Gepner point, these numbers are restricted to
be positive (or equivalently all negative). 
We claim that formula (\ref{genbeilchern}) above is the 
generalization of formula (\ref{beilchern}) valid for $\BP^n$.
Note that this in no way guarantees that the sheaf $E$ is stable or
semi-stable. That has to be checked separately as has been noted by
\cite{DD}.

Note that, as pointed out in \cite{dfr1}, the requirement that the $u_i$
be positive at the orbifold or Gepner point is a physical one. From all
known constructions for orbifolds it is clear that the D-brane spectrum
there does not separately produce brane-anti-brane bound states. Any
such bound state is already present in the spectrum. Hence there is no
need to add anti-branes by taking some of the $u_i$ to be negative as in
the theory at large volume. However this is not the case if we move away
from the orbifold point.

The dimension of the moduli space of the quiver gauge theory is given
by the formula\cite{dfr2}
\begin{equation}
d= 1- {1\over2} u^t\cdot C \cdot u
\end{equation}
where $C$ is the {\em Cartan matrix} associated with the quiver. We
claim that the natural Euler form associated with the quiver is
given by the (modified) Euler form associated with the helix. To be
precise, we claim that the Euler form associated with two
representations $u$ and $v$ of the Beilinson quiver is given by
\begin{equation}
\langle u, v\rangle = \sum_{ij} u_i\ v_j\ \chi(S_i,S_j)
\end{equation}
and the Cartan matrix is given by the symmetrization of the Euler
form
\begin{equation}
(u,v)= \langle u, v\rangle + \langle v, u\rangle 
\end{equation}
Hence, 
\begin{equation}
d= 1- {1\over2}(u,u)
\end{equation} 
We have checked this formula in some simple examples. However, 
the quiver gauge theory may have several branches and this formula
holds for a specific branch\footnote{We thank M. Douglas for an useful
communication in this regard. Details of these issues are to appear in
his forthcoming paper.}.

\subsection{Obtaining the McKay quiver}

So far we have been restricting our attention to what we have called
the generalized Beilinson quiver. However, in general at an orbifold
point, and 
presumably more generally,at a Landau-Ginzburg point, this cannot be the
full story. From the work of \cite{dfr2}, it is clear that the 
quiver associated to the orbifold point is the McKay quiver. In the case
of $\BP^n$, the McKay quiver is a closed quiver and the Beilinson quiver
is produced by truncating one of the links. Thus, in the $\BP^2$ case the
McKay quiver is a triangle with three vertices with three lines joining
each vertex in ordered fashion. This clearly reflects the $Z_3$ symmetry
of the orbifold point (see figure 3). 

How can we reconstruct the McKay quiver if we began with the Beilinson
quiver? We claim that the answer is clear from the construction of the GLSM
for the blow up of the $\BC^3/\BZ_3$ orbifold. While in the
large-volume limit, which is 
the setting for our construction of the $R_i$, the $S_i$
and the corresponding quiver, we have set the $p$-field to zero, this is
not true at the orbifold point. At the orbifold point, the $p$-field,
which has charge $-3$ in this case, acquires a non-trivial vacuum
expectation value, which gives rise to the twisted sectors of the
orbifold theory. The combinations $p\phi_i$, (for $i=1,2,3$)
where the $\phi_i$ are the
homogeneous coordinates of $\BP^2$, in fact can give rise to
homomorphisms from ${\cal O}(2)$ to ${\cal O}$ because the combinations
are of degree $-2$. This restores the missing link away from the
large-volume limit. 
We can implement this procedure in any of the theories that we have
considered. We give below the re-construction of the McKay-type quiver
for $\BP^2$ (figure 3) as well as for $\BP^{1,1,1,1,2}$ (figure 4). 
We depict the links obtained by using the $p$-field by dashed arrows.
\begin{figure}[ht]
\begin{center}
\leavevmode\epsfysize=3cm \epsfbox{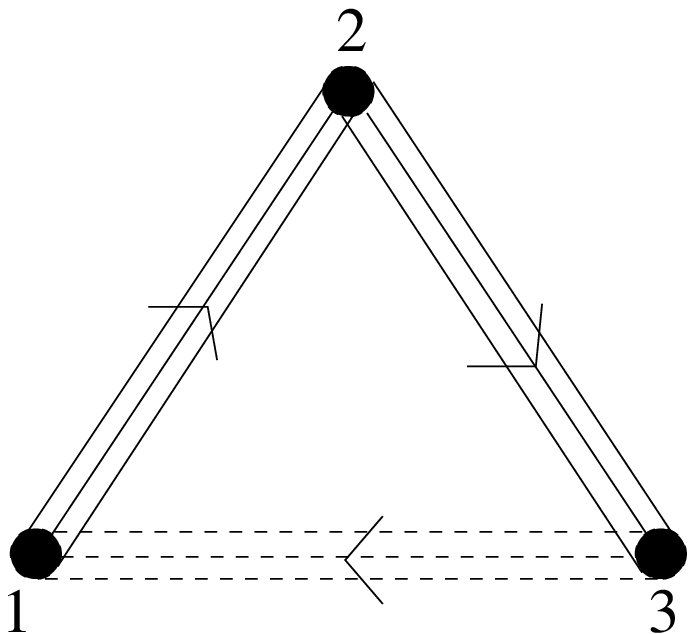}
\end{center}
\caption{McKay quiver for $\BP^2$}
\begin{center}
\leavevmode\epsfxsize=4cm \epsfbox{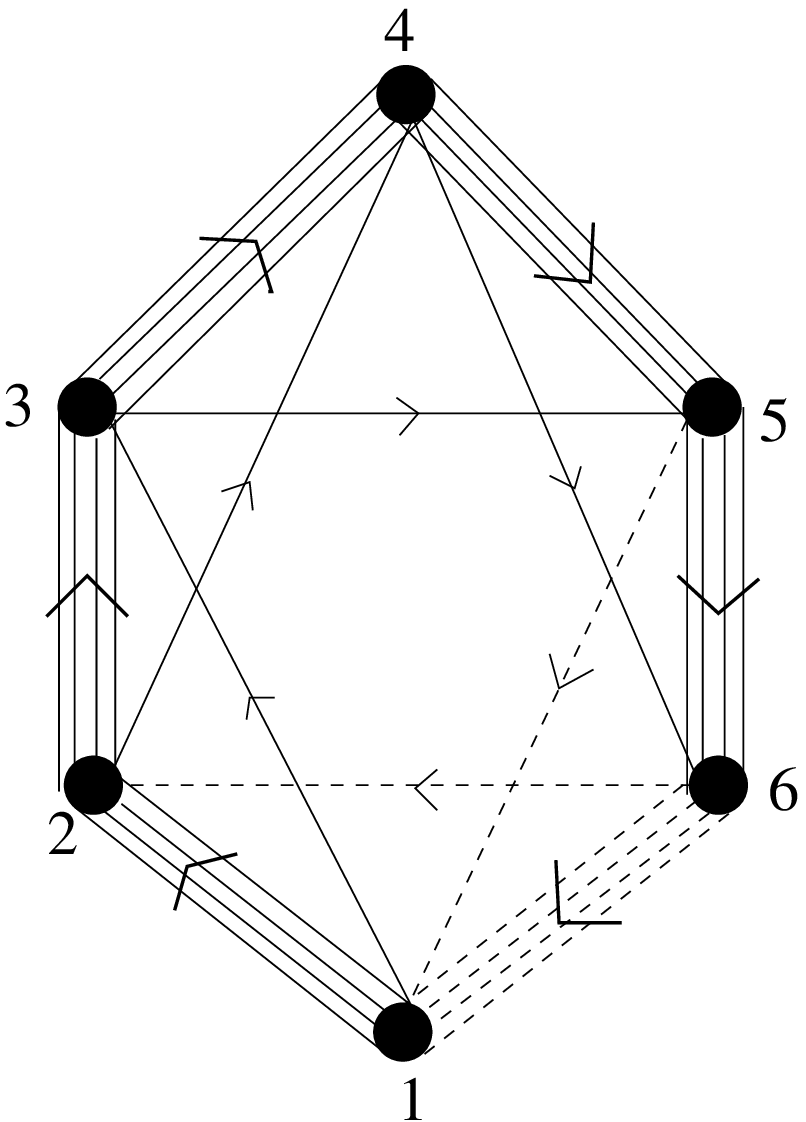}
\end{center}
\caption{McKay quiver for $\BP^{1,1,1,1,2}$}
\end{figure}
The quivers for the other examples that we consider can be worked out in a
similar fashion.

Note that this reconstruction of the McKay quiver from the Beilinson
quiver gives us non-trivial information. For instance, bound states that
use the link due to the $p$-field at the orbifold point are allowed in
the spectrum but are likely to decay before we reach the large-volume
limit, since there is no such link there and we have only the Beilinson
quiver. This point has also been made earlier, in \cite{dfr1}, but the
connection to the $p$-field was not realised there.

Another important example which illustrates the role of the $p$-field is
the flop transition as seen by the quiver gauge theory. Let us use the
local model for the flop as given in \cite{wittenphases} where there are
four fields, $(\phi_1, \ldots, \phi_4)$, with charges $(1,1,-1,-1)$.
In the limit of large and positive K\"ahler parameter $r$, 
the fields $\phi_3$ and $\phi_4$
are set to zero. These play the role of $p$-fields in this limit. The
corresponding quiver is given by the two nodes with two arrows joining
them as in the first of the diagrams in fig. 5. In the flopped phase,
where $r$ is large and negative, $\phi_1$ and $\phi_2$ are set to zero
and the other two are non-zero. The quiver for this phase is given by
reversing the two arrows as in the second diagram in fig. 5. 

\begin{figure}[ht]
\begin{center}
\leavevmode\epsfysize=4cm \epsfbox{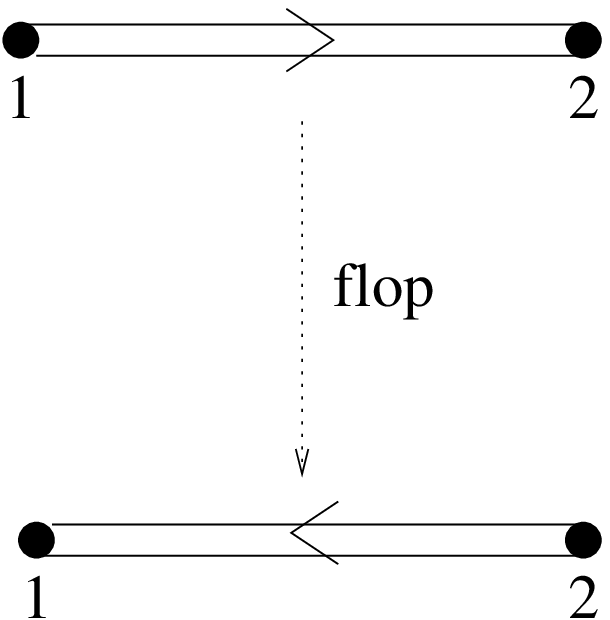}
\end{center}
\caption{Quivers for the flop transition}
\end{figure}

It is interesting to note that this operation, in the 
derived category ${\cal D}^b({\rm mod}-A)$ is equivalent to a change of
$t$-structure. In general, the addition and removal of arrows, as well
as the reversal of the direction of all arrows at a sink, under
appropriate circumstances, correspond to changes of $t$-structure.
We will not enter into a detailed discussion of the concept here but
refer the reader to \cite{partha} for definitions as well as other
details.  

This procedure, of using the $p$-fields to add links in the quiver
diagram appears to be completely general. In general at different
limit points corresponding to different ``phases'' of the CY manifold,
we can re-construct a quiver that would be in part, Beilinson-like, and
in part, McKay-like. Given that we can do this from our considerations
at the large volume end, we can thus obtain non-trivial information
about the D-brane spectrum at different limit points deep inside
different ``phases'' in the K\"ahler moduli space of CY manifolds. 
It would be interesting to apply this suggested technique to study 
the D-brane
spectrum at a limit point in the hybrid phase of a two-K\"ahler modulus
example and verify these, if possible, by an explicit computation using the
periods of the CY manifold at such a point. 

Note however that our procedure is not entirely complete. In general, on
a n-dimensional (weighted) projective space considered at the orbifold
or Gepner point, there are not only states in the vector representation
of $U(n)$ but also states in other, higher-dimensional representations.
We have not included these in our considerations so far, neither in the
Beilinson quiver nor in the corresponding McKay quiver. 

It is also time to sound a note of caution on 
one issue about which we have been somewhat
cavalier so far. All our considerations above apply to constructions in
the ambient variety $X$ in which the Calabi-Yau manifold is embedded as
a hypersurface $M$. Thus to recover objects on the CY manifold we need
to restrict these structures to the CY manifold. We have been careful
about this in the large-volume  limit but we have not really explored
this issue at the orbifold point. While this has not been a problem with
states associated to the $\sum_al_a=0$ orbit, this is unlikely to be true for
other states which appear as bound states of the exceptional ones. 
But this takes us to issues, especially those regarding bound states,on
which we will not have much to say in this paper.

It is worth emphasizing that the entire discussion here in sec. 5,
regarding quiver theories, is somewhat different in spirit
from the study of the McKay correspondence. Given an arbitrary orbifold,
it is not clear that there is a canonical resolution of the
singularities of that orbifold. However, here our orbifold-like points in
K\"ahler moduli space arise as the well-defined limit of a smooth CY
manifold and there is no issue at all about which resolution has to be
picked at the orbifold end of the story. But given that our aim here is
to study D-brane spectra at different points in the Kahler moduli space
of a CY manifold, this is good enough.

\section{Conclusion}

In this paper, we have presented a method of constructing exceptional
bundles on weighted projective spaces using the helix naturally
associated with the line bundle ${\cal O}$. These techniques generalise
in a rather straightforward manner to  spaces such as Grassmannians,
products of weighted projective spaces. However, in more general
situations such as these, the helix associated with ${\cal O}$ no longer
consists of only line bundles. For example, the foundation of the helix
on the Grassmannian $G(2,4)$ consists of spinor bundles and line
bundles\cite{karpov}. This is presumably related to non-abelian nature of 
the GLSM associated with the Grassmannian and will be an interesting
testing ground for some of conjectures  and may also require some
modifications before being adapted to more general situations. In this
paper, we have considered situations corresponding to CY compactifications
where the CY manifold appears as a hypersurface in weighted projective
space. There are other situations, such as F-theory 
compactifications where projective spaces and their blow-ups occur as
the base of an elliptic fibration where our methods may have direct
applications.

The special role played by the line bundle ${\cal O}$ is very much in
line with recent K-theoretic considerations based on Sen's study of
non-BPS branes\cite{sen}. For example, in IIB compactifications, all BPS branes
(as well as some non-BPS branes) can be obtained as bound states of a
certain number of D9 branes and anti-branes.
Further, D-brane charges in IIB theories
are classified by the K-theory group $K^0(X)$
where $X$ is the spacetime on which the string propagates\cite{wittenktheory}.
If we restrict our considerations to the case for $X=M\otimes\BR^{3,1}$,
where $M$ is a Calabi-Yau threefold and further, we consider only zero-branes
in the non-compact spacetime $\BR^{3,1}$, then the charges of the
particles (in IIA string theory)
will be classified by $K^0(M)$. Just as the D9-brane (and its antibrane)
generates all other branes in flat spacetime, it is natural to expect the 
six-brane (wrapping all of
$M$) to generate all lower branes and hence charges. As we have seen,
all exceptional branes can be generated by mutations. Other branes must
arise as bound states of these exceptional sheaves. This is in line with
the claim of \cite{dfr2} that the $\sum_a l_a\neq 0$
arise as bound states of the $\sum_a l_a=0$ states. These issues will be
discussed in a forthcoming publication.

One issue that we have not considered in this paper is the precise
location of lines of marginal stability. As we have pointed out
in the introduction itself, our analysis  allows us to identify
candidates for 
stable objects at special points in the K\"ahler moduli space. It will be
interesting to understand in more precise terms, where and how objects
decay.   It seems to us that the GLSM  would be the right setting to
study this issue further. Maybe, this might also be the way to obtain
the flow of gradings proposed by \cite{dfr1,dougtalk}. 

In a companion paper\cite{glsmnew}, we discuss in detail the explicit 
construction of
the exceptional bundles and some of their bound states in the GLSM with
boundary. These involve techniques similar to the constructions
which appear in $(0,2)$ compactifications of the heterotic string
though there are subtle differences.
We suitably modify this construction to describe $(0,2)$ multiplets
on the boundary of the worldsheet. As in our earlier paper\cite{glsm},
an important issue is the inclusion of an appropriate contact term
in order to have a sensible NLSM limit to the GLSM. We find that this
construction allows us to describe the tensoring of vector bundles by
suitable line bundles under large volume monodromy transformations.

\bigskip
\noindent {\bf Acknowledgments} We thank J. Biswas, K. Paranjape, 
R. Parthasarathy (Tata Institute) and S.  Ramanan for their patience in
answering our many queries and illuminating discussions.  
S.G. is supported in part by the Department of Science
and Technology, India under the grant SP/S2/E-03/96.

\appendix
\section{Mutations and Diagonalisation of Upper Triangular matrices}

Consider a helix ${\cal R}=(R_1,\cdots,R_n)$ of period $n$. The
intersection matrix $\chi({\cal R},{\cal R})$ is upper-triangular
with ones on the diagonal. We shall now prove that the process of
left mutations corresponds to the diagonalisation of the upper
triangular matrix $I_{ij}\equiv\chi({\cal R},{\cal R})$. 
We will define the following sets of bundles ${\cal R}^{(p)}$,
which are generated from the original helix ${\cal R}$. 
Note that the sets ${\cal R}^{(p)}$ for $p\neq 0,(n-1)$ are not
helices.  Consider
$$
{\cal R}^{(1)} = (R_1,L_1(R_2),L_2(R_3),\ldots,L_{n-1}(R_{n})) 
$$
and consider the matrix $I_{ij}^{(1)}\equiv\chi({\cal R},{\cal R}^{(p)})$. 
On using the formula for the Chern character of the mutated
bundle\footnote{We have, for simplicity, assumed that the map
$Ext^0(E_i,E_j)\otimes E_i\rightarrow E_j$ is injective. The
other case will include an additional minus sign, which can be
compensated for in the definition of the entries in ${\cal R}^{(p)}$
and is not important.}
$ {\rm ch}(L_{E_i}(E_j))={\rm ch}(E_j) - \chi(E_i,E_j) {\rm ch}(E_i)$, one can
see that the matrix $I_{ij}^{(1)}$ is also upper-triangular with ones
in the diagonal. In addition, $I_{i,i+1}^{(1)}=0$. The next set of
left mutations will  preserve the upper-triangular nature and will
have $I_{i,i+1}^{(2)}=I_{i,i+2}^{(2)}=0$. This is given by 
$$
{\cal R}^{(2)} = (R_1,L_1(R_2),L_1L_2(R_3),\ldots,L_{n-2}L_{n-1}(R_{n}))
$$
Notice, that the first two terms of ${\cal R}^{(2)}$ and ${\cal
R}^{(1)}$ are identical and in general, the first $(p-1)$ terms of
the collections ${\cal R}^{(p)}$ and ${\cal R}^{(p-1)}$  coincide.
The process ends after $(n-1)$ steps, after which the matrix $I^{(n)}_{ij}$
is the identity matrix.

Thus, the sequence  of left mutations gives rise to ${\cal S}\equiv
{\cal R}^{(n)}$ with the property
$$
\chi({\cal R},{\cal S}) = {\rm the\ identity\ matrix}
$$



\begin{thebibliography}{99}
\bibitem{RS}
A.~Recknagel and V.~Schomerus, ``D-branes in Gepner Models,'' 
Nucl. Phys. {\bf B531} (1998) 185, {\tt hep-th/9712186}.
\bibitem{quintic} I.~Brunner, M.~R.~Douglas, A.~Lawrence and C.
R\"omelsberger, ``D-branes on the Quintic,'' JHEP {\bf 0008} (2000) 015,
{\tt hep-th/9906200}.
\bibitem{diacgom} D.-E.~Diaconescu and J.~Gomis, ``Fractional Branes and
Boundary States in Orbifold Theories, JHEP {\bf 0010} (2000) 001, 
{\tt hep-th/9906242}.
\bibitem{dgepner}
M.~Gutperle and Y.~Satoh, ``D-branes in Gepner models
and supersymmetry," Nucl. Phys. {\bf B543} (1999) 73, {\tt
hepth/9808080}. \\
D.-E.~Diaconescu and C.~R\"omelsberger,
``D-Branes and Bundles on Elliptic Fibration,'' Nucl. Phys. {\bf B574} (2000)
245-262, {\tt hepth/9910172}.  \\
P.~Kaste, W.~Lerche, C.A.~Lutken and J.~Walcher, ``D-branes on K3 -
Fibrations,'' {\tt hepth/9912147} \\
M. Naka and M. Nozaki, ``Boundary states in Gep\-ner mod\-els,''
{\tt hepth/0001037}. \\
I.~Brunner and V.~Schomerus, ``D-branes at Singular Curves of
Calabi-Yau Compactifications,'' {\tt hepth/0001132}.\\
J.~Fuchs, C.~Schweigert, J.~Walcher, ``Projections in string theory
and boundary states for Gepner models,'' {\tt hep-th/0003298}.
\bibitem{stt}
S.~Govindarajan, T.~Jayaraman and T.~Sarkar, ``Worldsheet approaches to
D-branes on supersymmetric cycles,'' Nucl. Phys. {\bf B580} (2000) 519-547, 
{\tt hep-th/9907131}. \\
S. Govindarajan and T. Jayaraman, ``On the
Landau-Ginzburg description of Boundary CFTs and special Lagrangian 
submanifolds, JHEP {\bf 07} (2000) 016, {\tt hep-th/0003242}.
\bibitem{scheid} E.~Scheidegger, ``D-branes on some one- and two-parameter
Calabi-Yau hypersurfaces,'' JHEP {\bf 0004} (2000) 003, {\tt hepth/9912188}.
\bibitem{wittenphases}
E.~Witten, ``Phases of $N=2$ Theories in Two Dimensions,''
Nucl. Phys. {\bf B403} (1993) 159, {\tt hep-th/9301042}.
\bibitem{topics}
M.R.~Douglas, ``Topics in D-Geometry,''
Class. Quant. Grav. {\bf 17} (2000) 1057, {\tt hep-th/9910170}.
\bibitem{dfr1}
M.~R.~Douglas, B.~Fiol and C.~R\"omelsberger,
``Stability and BPS branes,'' {\tt hep-th/0002037}.
\bibitem{dfr2} M.~R.~Douglas, B. Fiol and C.~R\"omelsberger, ``The spectrum
of a non-compact Calabi-Yau manifold,'' {\tt hep-th/0003263}.
\bibitem{DD} M.~R.~Douglas and D.-E..~Diaconescu, ``D-branes on Stringy
Calabi-Yau Manifolds,'' {\tt hep-th/0006224}.
\bibitem{dougtalk}
M.~R.~Douglas, ``D-branes on Calabi-Yau manifolds,''
{\tt math.ag/0009209}.
\bibitem{dgm}
M.~R.~Douglas, B.~Greene, D.~Morrison, ``Orbifold Resolution by D-branes,''
Nucl. Phys. {\bf B506} (1997) 84, {\tt hep-th/9704151}. 
\bibitem{inverse} 
T.~Muto, ``D-branes on orbifolds and topology change,''
Nucl.\ Phys.\  {\bf B521} (1998) 183, {\tt hep-th/9711090}. \\
B.~Greene, `` D-brane topology changing transitions,''
Nuc. Phys. {\bf B525} (1998) 284, {\tt hep-th/9711124}. \\
S.~Mukhopadhyay and K.~Ray, ``Conifolds from D-branes,''
Phys.\ Lett.\  {\bf B423} (1998) 247, {\tt hep-th/9711131}. \\
B.~Feng, A.~Hanany and Y.~He,
``D-brane gauge theories from toric singularities and toric duality,''
{\tt hep-th/0003085}. \\
T.~Sarkar, ``D-brane gauge theories from toric singularities of the form
$\BC^3/\Gamma$ and $\BC^4/\Gamma$,'' {\tt hep-th/0005166}.
\bibitem{Ito-Nak} Y.~Ito and H.~Nakajima, `` McKay correspondence and
Hilbert schemes in dimension three,'' {\tt math.AG/9803120}.
\bibitem{reid1} Miles Reid, `` McKay correspondence,'' {\tt
alg-geom/9702016}.
\bibitem{reid2} Miles Reid, `` La correspondance de McKay,'' S\'eminaire
Bourbaki, 52\`eme ann\'ee, Novembre 1999, no. 867, {\tt
math.AG/9911165}, to appear in Ast\'erisque 2000.
\bibitem{glsm} S.~Govindarajan, T.~Jayaraman and T.~Sarkar, ``On D-branes
from Gauged Linear Sigma Models,'' {\tt hep-th/0007075}, to appear in
Nuc. Phys. B. 
\bibitem{HIV} K.~Hori, A.~Iqbal and C.~Vafa, ``D - Branes And Mirror
Symmetry,'' {\tt hep-th/0005247}.
\bibitem{rudakov} {\em Helices and Vector Bundles: Seminaire
Rudakov}, A.N. Rudakov et. al., Lond. Math. Soc. Lec. Notes Series
148, Cambridge Univ. Press (1990).
\bibitem{zaslow} E.~Zaslow, ``Solitons and Helices: the search for a
math-physics bridge,'' Comm. Math. Phys. {\bf 175} (1996) 337,
{\tt hep-th/9408133}.
\bibitem{glsmnew} S. Govindarajan and T. Jayaraman (to appear)
\bibitem{lty} B.H.~Lian, A.~Todorov, S.-T.~Yau, ``Maximal
Unipotent Monodromy for Complete Intersection CY Manifolds,'' {\tt
math/0008061}.
\bibitem{kt} A.~Klemm and S.~Theisen, `` Considerations of one-modulus
Calabi-Yau compactifications: Picard-Fuchs equations, K\"ahler potentials
and mirror maps,'' Nucl. Phys. {\bf B389} (1993) 153-180 {\tt hep-th/9205042} 
\bibitem{font} A. Font, ``Periods and Duality Symmetries in
Calabi-Yau Compactifications,'' Nucl. Phys. {\bf B391} (1993) 358, {\tt
hep-th/9203084}.
\bibitem{twopar} 
P.~Candelas, X.~de la Ossa, A.~Font, S.~Katz and  D.~R.~Morrison, 
``Mirror Symmetry for Two Parameter Models -- I,'' 
Nucl. Phys. {\bf B416} (1994) 481-538, {\tt hep-th/9308083} \\
P.~Candelas, A.~Font, S.~Katz and  D.~R.~Morrison, 
``Mirror Symmetry for Two Parameter Models -- II,'' 
Nucl. Phys. {\bf B429} (1994) 626-674, {\tt hep-th/9403187} \\
S.~Hosono, A.~Klemm and S.~Theisen, ``Mirror Symmetry, Mirror Map and
Applications to Calabi-Yau Hypersurfaces,'' Commun. Math. Phys. {\bf
167} (1995) 301-350, {\tt hep-th/9308122} 
\bibitem{spgeom} A. Ceresole, R. D'Auria, S. Ferrara, W. Lerche, J.
Louis, ``Picard-Fuchs Equations and Special Geometry,''
Int. J. Mod. Phys. {\bf A8} (1993) 79-114, {\tt hep-th/9204035}. \\
S. Hosono, A. Klemm, S. Theisen and S.-T. Yau, ``Mirror Symmetry,
Mirror Map and Applications to Complete Intersection Calabi-Yau
Spaces,'' Nucl. Phys. {\bf B433} (1995) 501-554, {\tt hep-th/9406055}.
\bibitem{bpsalgebra} J.~A.~Harvey and G.~Moore, ``On the algebra of BPS
states,'' Commun. Math. Phys. {\bf 197} (1998) 489-519, {\tt hep-th/9609017}.
\bibitem{bondal} A.I. Bondal, ``Helixes, Representations of Quivers
and Koszul Algebras,'' in ref. \cite{rudakov}.
\bibitem{beilinpn} A.~A.~Beilinson, ``Coherent sheaves on $\BP^n$ and
problems of linear algebra,'' Funct. Anal. Appl. {\bf 12}
(1978) 214-216.
\bibitem{gorrud} A.~L.~Gorodentsev and A.~N.~Rudakov, ``Exceptional
Bundles on  Projective Space,'' Duke Math. J. {\bf 54} (1987) 115-130.
\bibitem{drezet} J.-M. Dr\'ezet, ``Espaces abstraits de morphismes et
mutations,'' {\tt alg-geom/9603016}.
\bibitem{bondal1} A.~.I.~Bondal, ``Representation of associative
algebras and coherent sheaves,'' Math. USSR-Izv. {\bf 34} (1990)
23-42.
\bibitem{partha} R.~Parthasarathy, ``t-structures in the derived
category of representations of quivers,'' Proc. Indian Acad. Sci. (Math.
Sci.) {\bf 98} (1988) 187-214.
\bibitem{karpov} B.~V.~Karpov, ``A symmetric helix on the Pl\"ucker
quadric,'' in \cite{rudakov}.
\bibitem{sen}
A.~Sen,
``Stable non-BPS states in string theory,''
JHEP {\bf 9806}, 007 (1998)
{\tt hep-th/9803194}. \\
``Stable non-BPS bound states of BPS D-branes,''   
JHEP {\bf 9808}, 010 (1998)
{\tt hep-th/9805019}. \\
A.~Sen,
``Tachyon condensation on the brane antibrane system,''
JHEP {\bf 9808}, 012 (1998)
{\tt hep-th/9805170}.
\bibitem{wittenktheory} E.~Witten, ``D - Branes And K - Theory,'' JHEP
{\bf 9812} (1998) 019, {\tt hep-th/9810188}.
\end{thebibliography}
\end{document}